\documentstyle[epsf,epsfig,12pt]{article}  
\begin{document}

\newcommand{\be}{\begin{equation}}

\begin{titlepage}

\pagenumbering{arabic}
\vspace*{-1.cm}
\begin{tabular*}{15.cm}{l@{\extracolsep{\fill}}r}
&
\hfill \bf{DEMO-HEP 97/09}
\\
& 
\hfill 20 November, 1997
\\
\end{tabular*}
\vspace*{2.cm}
\begin{center}
\Large 
{\bf Accurate Estimation of  the Trilinear Gauge Couplings Using
Optimal Observables Including Detector Effects}

\vspace*{2.cm}
\normalsize { 
{\bf G. K. Fanourakis, D. Fassouliotis and S. E. Tzamarias} \\
\vspace*{0.3cm}
{\footnotesize N.C.S.R. Demokritos}\\
}
\end{center}
\vspace{\fill}
\begin{abstract}
\noindent

This paper describes the definition of maximum likelihood equivalent 
estimators of the Trilinear Gauge Couplings
which include detector effects. The asymptotic properties of these estimators
as well as their unbiasedness and efficiency when dealing with finite 
statistical samples are demonstrated by Monte Carlo experimentation, using simulated 
events corresponding to the reaction $e^+e^-\rightarrow q\bar{q}l\nu$ at 172 GeV.
Emphasis is given to the determination
of the expected efficiencies in extracting the $\alpha_{W\phi}$, 
$\alpha_{W}$ and $\alpha_{B\phi}$ couplings from LEPII data, which 
in this particular case found to be close to the maximum possible.

\end{abstract}
\vspace{\fill}

\end{titlepage}

\pagebreak

\begin{titlepage}
\mbox{}
\end{titlepage}

\pagebreak

\setcounter{page}{1}    


\section{Introduction}
It has been often emphasized that the likelihood estimators can reach the 
lower bound of the Cramer-Rao \cite{eadie} inequality,
\begin{equation}
V(\hat{\lambda}) \geq 
[-E(\frac{ \partial^{2} \ln{L} } 
{\partial \lambda^{2} } )_{\lambda = \hat{\lambda} }]^{-1}
\geq \frac{-1} {{\cal{I}}_{\lambda} }   \label{eq:1}
\end{equation}
where:
\begin{description}
\item[$\hat{\lambda}$] is the likelihood estimation of a parameter $\lambda$
\item[$L$] is the likelihood function
\item[$V$] and $E$ are the variance and expected value operators respectively
\item[$\cal{I}_{\lambda}$] denotes the maximum available information with 
respect to $\lambda$.
\end{description}
The right hand side bound depends on the construction of the probability density function
(p.d.f.). Thus, as has been shown in \cite{sekulin} and \cite{yellow}, 
in order to reach the maximum accuracy in
estimating the Trilinear Gauge Couplings, the complete 
kinematical information has to be included in $L$. However, when the measured
kinematic vectors are used, the experimental uncertainties must be 
explicitly taken into account
in the p.d.f. formulation. In other words, the data must be compared with the 
distribution:
\begin{equation}
P(\vec{\Omega} ;\vec{\lambda} )=\int g(\vec{V};\vec{\lambda})\cdot R(\vec{V},\vec{\Omega})d\vec{V}
\label{eq:2}
\end{equation}
rather than with 
\begin{equation}
g(\vec{V};\vec{\lambda})=\frac{(d\sigma(\vec{V};\vec{\lambda})/d\vec{V})\cdot 
\epsilon (\vec{V})}{\sigma_{obs.}(\vec{\lambda})}
\label{eq:3}
\end{equation}
where
\begin{description}
\item $\vec{V} = \{V_{1},\ldots ,V_{m}\}$  and 
$\vec{\Omega} = \{\Omega_{1}, \ldots, \Omega_{m}\}$
 are the true and measured m-dimensional 
kinematic vectors respectively.
\item[$\vec{\lambda}$] is the set of $\rho$ physics parameters 
$\{ \lambda_{1}, \ldots ,\lambda_{\rho} \}$
which are needed to define completely the p.d.f.
\item[$\epsilon(\vec{V})$] is the selection efficiency function taking values 
continuously between zero and one,
\item[$d\sigma(\vec{V};\vec{\lambda})/d\vec{V}$] is the differential cross 
section.   
\item[$\sigma_{obs.}(\vec{\lambda})$] is the total observed cross 
section defined as\footnote{Throughout this paper the integrations are meant to be over
the whole phase space.}:
\begin{equation}
 \sigma_{obs.}(\vec{\lambda})= \int(d\sigma(\vec{V};\vec{\lambda})/d\vec{V})\cdot 
\epsilon (\vec{V})d\vec{V}
\label{eq:4}
\end{equation}
\item[$R(\vec{V},\vec{\Omega})$] is the resolution function, or else the 
probability  for the true
kinematic vector $\vec{V}$ of an event to be measured as $\vec{\Omega}$.
\item[$g(\vec{V};\vec{\lambda})$] is the p.d.f. according to which the 
vectors $\vec{V}$ are distributed and
\item[$P(\vec{\Omega} ;\vec{\lambda})$] is the p.d.f. according to which the 
vectors $\vec{\Omega}$ are distributed.
\end{description}

The use of (\ref{eq:2}), in multi-dimensional phase space, faces several 
difficulties of which the parameterization of the resolution functions 
is the most serious. The 
often offered remedies consist of either using  (\ref{eq:3}) instead of 
(\ref{eq:2}) with the 
hope that possible biases will be determined afterwards by Monte Carlo (M.C.) 
experimentation
or of integrating over some of the kinematic variables and 
paramererizing the resolution
with respect to the remaining fewer kinematic components.

This paper concentrates on the definition of likelihood equivalent estimators
which, whilst including the description of all the detector
effects, retain the maximum likelihood sensitivity in extracting the values 
of the Trilinear Gauge Couplings. The next Section  briefly describes
the topological and kinematic features of the data samples used in this analysis. In Section 3, ways of
reducing  the number of the necessary kinematic variables are discussed when the detector 
effects are negligible. In Section 4 the results obtained in the ideal case are extrapolated to the case
when the detector effects are important. A demonstration of the optimal 
properties of the proposed techniques
is given at Section 5 by  Monte Carlo experimentation. Finally, Section 6 
compiles the conclusions of this analysis.

\section{The Four-Fermion Semileptonic Final States}

One of the important physics issues of the LEPII program is the determination of 
the non-Abelian self couplings ( Trilinear Gauge Couplings, in the following TGC's) of the
carriers of the electroweak interaction. The measurement of the WWV couplings, where W
stands for the W gauge boson whilst V denotes the $\gamma$ or $Z^{0}$ neutral carriers,
is possible via WW production \cite{yellow}  through diagrams involving exchange of
$Z^{0}$ and $\gamma$. Limiting the phenomenological analysis to the 
contribution of gauge invariant operators of dimension less than six 
\cite{gounar}, there are 3 CP conserving ($\alpha_{W\phi}$, 
$\alpha_{W}$ and $\alpha_{B\phi}$) and 2 CP violating ($\tilde{\alpha}_{BW}$
and $\tilde{\alpha}_{W}$)  couplings which by deviating from their Standard 
Model values, could point to the existence of new physics.

Several final state topologies, corresponding to the production of a pair or
of a single W and to the different W decay modes,
can be used to determine the TGC's. In this paper only final states including a charged lepton,
two hadronic jets and an invisible neutrino are considered. These final states, in the following
semileptonic final states, suffer less from kinematic ambiguities and consequently a
more precise measurement can be made.

However, W pair production diagrams alone are not sufficient to express the 
production dynamics of the semileptonic states \cite{yellow}.  
Throughout this paper full four-fermion phenomenological
models including Coulomb corrections and Initial State Radiation (ISR)
\cite{erato},\cite{exca} are used to describe the kinematic distributions and/or
to produce Monte Carlo events as functions 
of the CP conserving couplings.

The detector simulation program DELSIM \cite{delsim} was used to study the 
effect of the distortion of the kinematic distributions, due to the event 
selection efficiency and the detector resolution, on
the measurement of the TGC's. Special care was taken in selecting the 
semileptonic four fermion final state
topologies in order to avoid contamination from products irrelevant to the determination of the TGCs. 
The selection criteria,  based on the lepton identification jet reconstruction
and missing visible energy, are discussed in details in \cite{jeru}. An improvement in
the measurement accuracy of the kinematic vectors was achieved by a six-constraint kinematic fit.
\footnote{ In the
6c kinematical fit the total four-momentum vector was constrained to (0,0,0,172 $GeV$)
and the invariant masses of the hadronic and the leptonic system were required to be equal
to 80.35 $GeV/c^{2}$.}

After the constrained  fit, all the eight kinematic variables needed to 
specify an event completely were determined up to a degeneracy in the charge 
of the hadronic jets.
The latter resulted in a loss of information by averaging  
the squared matrix elements corresponding to the two momentum assignments  
\cite{sekulin} \cite{yellow}.

\section{The Ideal Case}
The differential cross section
is parameterized as a quadratic function of the TGC's \cite{yellow} 
of the form:
\begin{equation}
d\sigma(\vec{V};\vec{\lambda})/d\vec{V} = c_{0}(\vec{V}) +
\sum_{k}c^{k}_{1}(\vec{V})\cdot\lambda _{k} + 
\sum_{\mu ,k}c_{2}^{\mu k}(\vec{V})\cdot \lambda_{\mu}\cdot\lambda_{k}
\label{eq:4b}
\end{equation}
where the vector $\vec{V}$ is of dimension eight and
$\vec{\lambda} = \{ \lambda_{1}, \ldots ,\lambda_{\rho} \}$ is the set 
of the TGC's which are allowed to deviate
from their Standard Model (S.M.) values.
Following eq. (\ref{eq:4}), the total cross section can be expressed as:
\begin{equation} 
\sigma_{tot}(\vec{\lambda} )=  
S_{0}+\sum_{k} S_{1}^{k}\cdot \lambda_{k}+ \sum_{\mu k} S_{2}^{\mu k}\cdot 
\lambda_{\mu}
\cdot\lambda_{k}
\label{eq:5}
\end{equation}
\begin{equation} 
S_{i}^{j} = \int c_{i}^{j}(\vec{V})d\vec{V}
\label{eq:6}
\end{equation}

When the detector effects are negligible the p.d.f. takes the form:
\begin{equation}
g(\vec{V},\vec{\lambda}) =
 \frac{c_{0}(\vec{V}) +\sum_{k} c^{k}_{1}(\vec{V}) \cdot\lambda _{k} + 
\sum_{\mu ,k} c_{2}^{\mu k}(\vec{V}) \cdot\lambda_{\mu} \cdot\lambda_{k} }
{S_{0}+\sum_{k} S_{1}^{k}\cdot\lambda_{k}+\sum_{\mu k} S_{2}^{\mu k}\cdot\lambda_{\mu}
\cdot\lambda_{k}}
\label{eq:7}
\end{equation}
and the likelihood function 
for a set of N observed events with kinematic vectors $\vec{V}_{n}$, $n=1, \ldots ,N$
is defined as:
\begin{equation}
L(\vec{V}_{1}, \ldots ,\vec{V}_{N};\vec{\lambda}) = \prod_{n=1}^{N}g(\vec{V}_{n},\vec{\lambda})
\label{eq:8}
\end{equation}

\subsection{Optimal Variables}
 The kinematic distribution with respect to a subset of the components of $\vec{V}$, let's say
$X=\{V_{1}, \ldots ,V_{\nu}\}$ with $\nu <m$, is found by integrating (\ref{eq:7}) as:
\begin{equation}
\varpi(q_{1}, \ldots ,q_{\nu};\vec{\lambda} ) = \int g(\vec{V};\vec{\lambda})\cdot 
\delta (q_{1}-V_{1}) \cdots  \delta (q_{\nu }-V_{\nu })dV_{1} \cdots dV_{\nu} \cdots dV_{m}
\label{eq:9}
\end{equation}
and the corresponding likelihood function for N events is:
\begin{equation}
L(X_{1}, \ldots ,X_{N};\vec{\lambda}) =\prod_{n=1}^{N}\varpi(X_{n};\vec{\lambda} )
\label{eq:10}
\end{equation}
The likelihood estimators of the couplings $\vec{\lambda}$ are defined as the 
solutions of the following system of equations :
\begin{equation}
\int [\frac{\partial \ln{L(X_{1}, \ldots ,X_{N};\vec{\lambda})}}{\partial \lambda_{j}}
]_{\vec{\lambda} = \hat{\lambda}} \cdot L(X_{1}, \ldots ,X_{N};\vec{\lambda} = \vec{\lambda}_{true})dX_{1} \cdots dX_{N}
= 0
\label{eq:11}
\end{equation}
or equivalently \cite{eadie}
\begin{equation}
N\cdot\int [\frac{\partial \ln{\varpi(X;\vec{\lambda})}}{\partial \lambda_{j}}
]_{\vec{\lambda} = \hat{\lambda}} \cdot \varpi(X;\vec{\lambda} = \vec{\lambda}_{true})dX
= 0
\label{eq:11b}
\end{equation}
where $\vec{\lambda}_{true}$, $\hat{\lambda}$
 are the true and the estimated values of the
couplings respectively.
If a single set of N observed events is available eq. (\ref{eq:11b}) is 
approximated as:
\begin{equation}
\frac{\partial \ln{L(X_{1}, \ldots ,X_{N};\vec{\lambda})}}{\partial 
\lambda_{j}}]_{\vec{\lambda} = \hat{\lambda}}
= \sum_{k=1}^{N}\frac{\partial \ln{\varpi(X_{k};\vec{\lambda})}}{\partial 
\lambda_{j}}]_{\vec{\lambda} = \hat{\lambda}} = 0
\label{eq:12}
\end{equation}

The maximum likelihood estimation  (\ref{eq:12}) 
satisfies at the asymptotic limit the left side bound of (\ref{eq:1}). However, 
the right side bound of
the Cramer-Rao inequality is not necessarily satisfied after projecting the 
p.d.f.
Indeed, in \cite{sekulin} and \cite{yellow} traditional kinematic distributions 
such as the W production angle, the 
charged lepton angle with respect to the W direction in the W rest frame etc.
have been rated 
according to their efficiency in estimating  the TGC's and they have been
found inferior to the eight-fold unbinned likelihood fit. 

Searching for a subset of the kinematic components which contain all
the available information
with respect to $\vec{\lambda}$,  one
has to start from the error matrix of  the likelihood estimators  when 
all the kinematic variables have been used, as in (\ref{eq:7}) and 
(\ref{eq:8}). In the asymptotic limit
this matrix has elements given by (\ref{eq:1}), i.e.:
\begin{eqnarray}
V(\lambda_{i},\lambda _{j}) 
& = & [-E(\frac{ \partial^{2} \ln{L(\vec{V}_{1}, 
\ldots ,\vec{V}_{N};\vec{\lambda}=\hat{ \lambda})}}
{\partial \lambda_{i} \partial \lambda_{j}})]^{-1} \nonumber \\
& = & \frac{-1}{N} \cdot [ \int \frac{ \partial^{2} \ln{ g(\vec{V};\vec{\lambda}
=\hat{ \lambda} ) } }{\partial \lambda_{i} \partial \lambda_{j}}
g(\vec{V};\vec{\lambda}= \vec{\lambda}_{true})d\vec{V}]^{-1} 
\label{eq:13}  
\end{eqnarray}
The second derivative in (\ref{eq:13}) depends on the kinematic vectors 
$\vec{V}$ through terms such as
 $c_{1}^{k}(\vec{V})/c_{0}(\vec{V})$ and $c_{2}^{\mu k}
(\vec{V})/c_{0}(\vec{V})$.
This has significant  implications especially when only one 
parameter TGC models are used. In this case the p.d.f. is written as:
\begin{eqnarray}
g(\vec{V};\lambda ) = \frac{c_{0}(\vec{V})+c_{1}(\vec{V})\cdot\lambda+c_{2}(\vec{V})\cdot\lambda^{2}}
{S_{0}+S_{1}\cdot\lambda+S_{2}\cdot\lambda^{2}}
\label{eq:14}
\end{eqnarray}
where the terms $c_{0}(\vec{V})$, $c_{1}(\vec{V})$ and $c_{2}(\vec{V})$ correspond
to the particular choice of the TGC $\lambda$.\\
The error in the maximum likelihood estimation of $\hat{\lambda}$ will be:
\begin{eqnarray}
V(\hat{\lambda}) =&&    \frac{-1}{N}\cdot [ 
                   -\frac{2\cdot S_{2}\cdot(S_0+S_{1}\cdot\hat{\lambda}+
S_{2}\cdot\hat{\lambda}^{2})-(S_{1}+2\cdot S_{2}\cdot\hat{\lambda})^{2}}
{(S_{0}+S_{1}\cdot\hat{\lambda}+S_{2}\cdot\hat{\lambda}^{2})^{2}}  
\nonumber \\ 
                && +\int \frac {2\cdot Q_{2}(\vec{V})\cdot(1+Q_{1}
(\vec{V})\cdot \hat{\lambda}+Q_{2}(\vec{V})\cdot\hat{\lambda}^{2})
-(Q_{1}(\vec{V})+2\cdot Q_{2}(\vec{V})\cdot\hat{\lambda})^{2}}
{(1+Q_{1}(\vec{V})\cdot\hat{\lambda}+Q_{2}(\vec{V})\cdot\hat{\lambda}^{2})^{2}} \nonumber \\  
&&\cdot g(\vec{V};\lambda_{true})d\vec{V} ]^{-1} 
\label{eq:15} 
\end{eqnarray}
where
\begin{eqnarray}
 Q_{1}(\vec{V})  =  \frac{c_{1}(\vec{V})}{c_{0}(\vec{V})}
\label{eq:15q1}
\end{eqnarray}
\begin{eqnarray}
 Q_{2}(\vec{V})  =  \frac{c_{2}(\vec{V})}{c_{0}(\vec{V})}
\label{eq:15q2}
\end{eqnarray}

However, when the likelihood function is expressed in terms of the  
projected probability distribution: 
\begin{eqnarray}
{ \varpi}(q_{1},q_{2};\lambda) =\int g(\vec{V};\lambda )
\cdot\delta(q_{1}-Q_{1}(\vec{V}))
\cdot\delta(q_{2}-Q_{2}(\vec{V}))d\vec{V}
\label{eq:16}
\end{eqnarray}
the variance of the estimated coupling will be:
\begin{eqnarray}
 V(\hat{\lambda})  =&&  \frac{-1}{N}\cdot [ 
 -\frac { 2\cdot S_{2} \cdot (S_{0}+S_{1} \cdot \hat{\lambda}+S_{2} 
 \cdot \hat {\lambda}^{2}) - (S_{1}+2 \cdot S_{2} \cdot \hat{\lambda})^{2}}
 {(S_{0}+S_{1} \cdot \hat{\lambda}+ S_{2} 
 \cdot \hat{\lambda}^{2})^{2} }  
\nonumber \\
&&+\int \int \int\{ \frac 
 {2\cdot q_{2} \cdot (1+q_{1} 
 \cdot \hat{\lambda } +q_{2} \cdot \hat{\lambda}^{2}) - ( q_{1}+ 2\cdot q_{2} 
 \cdot \hat{\lambda} )^{2} } { (1 + q_{1} \cdot \hat{\lambda} +q_{2} 
 \cdot \hat{\lambda}^{2} )^{2} } \nonumber \\ 
&& \cdot  g(\vec{V}; \lambda_{true} ) \cdot \delta ( q_{1}- Q_{1} (\vec{V}))
 \cdot \delta (q_{2}-Q_{2}(\vec{V}))\} d\vec{V}dq_{1} dq_{2} ]^{-1} 
\label{eq:17}
\end{eqnarray}

Reversing the order of integration in (\ref{eq:17})
and integrating first with respect to $q_{1}$ and $q_{2}$ it is easy to see 
that the
achieved accuracy by employing the projected distribution (\ref{eq:16}) is 
the same as when the complete multi-dimensional distribution (\ref{eq:14}) is used.

\subsection{The Optimal Observables}
 The reduction of the number of necessary kinematic components is not
so dramatic when more than one TGC are to be simultaneously extracted from 
the data\footnote{ As an example, in a two couplings
model 
the eight kinematic variables can be  reduced  to five.}. There is, though,
the possibility of a further reduction in the number of the necessary 
kinematic variables
by expanding the p.d.f. in a Taylor series and keeping only the linear terms. 
Returning
to the general case of estimating simultaneously $\rho$ couplings, the 
distribution function 
(\ref{eq:7}) is approximated in the neighborhood of the expansion point 
$\lambda^{0}$ as:

\begin{eqnarray}
g(\vec{V};\vec{\lambda}) \simeq \frac{y_{0}(\vec{V};\vec{\lambda}^{0})}{Z_{0}(\vec{\lambda}^{0})}\cdot [1+\sum_{k=1}^{\rho}
(\frac{y_{1}^{k}(\vec{V};\vec{\lambda}^{0})}
{y_{0}(\vec{V};\vec{\lambda}^{0})}-\frac{Z_{1}^{k}(\vec{\lambda}^{0})}{Z_{0}(\vec{\lambda}^{0})})\cdot\Delta_{k}]
\label{eq:19}
\end{eqnarray}

where 

\begin{equation}
 y_{0}(\vec{V};\vec{\lambda}^{0})=c_{0}(\vec{V})+\sum_{k} 
C_{1}^{k}(\vec{V})\cdot{\lambda}^{0}_{k}+
\sum_{\mu k} c_{2}^{\mu k}(\vec{V})\cdot{\lambda}^{0}_{\mu}\cdot{\lambda}^{0}_{k} 
\label{eq:20a} 
\end{equation}

\begin{eqnarray}
y_{1}^{k}(\vec{V};\vec{\lambda}^{0})  =  c_{1}^{k}(\vec{V})+\sum_{\mu } C_{2}^{\mu k}
(\vec{V})\cdot{\lambda}^{0}_{\mu} 
\label{eq:20b} 
\end{eqnarray}
\begin{equation}
Z_{\rho}^{k} =  \int y_{\rho}^{k}d\vec{V}
\end{equation}
 and $\Delta_{k}  = {\lambda}_{k}-{\lambda}^{0}_{k} $.

Furthermore, by projecting (\ref{eq:19}) as:

\begin{equation}
\varpi(O_{1}(\vec{\lambda}^{0}), \ldots ,O_{\rho}(\vec{\lambda}^{0});\vec{\lambda}) = 
\nonumber
\end{equation}
\begin{equation}
\int g(\vec{V};\vec{\lambda})\cdot\delta(O_{1}(\vec{\lambda}^{0})-\frac{y_{1}^{1}(\vec{V};\vec{\lambda}^{0})}
{y_{0}(\vec{V};\vec{\lambda}^{0})}) \cdots  \delta(O_{\rho}(\vec{\lambda}^{0})-\frac{y_{1}^{k}(\vec{V};\vec{\lambda}^{0})}
{y_{0}(\vec{V};\vec{\lambda}^{0})}) d \vec{V}
\label{eq:21}
\end{equation}
the number of the necessary kinematic variables are reduced to as many as
the number of parameters to be fitted simultaneously
without losing information. It has been also shown in \cite{opob} that there are simple statistics
equivalent to the maximum likelihood estimators. These are the mean 
values of 
$O_{k}(\vec{V},\vec{\lambda}^{0})$ ,  called Optimal Observables, defined as:
\begin{equation}
O_{k}(\vec{V};\vec{\lambda}^{0})=\frac{y_{1}^{k}
(\vec{V};\vec{\lambda}^{0})}{y_{0}(\vec{V};\vec{\lambda}^{0})}
\label{eq:24a}
\end{equation}
which are linearly related to $\hat{\Delta_{i}} ( = \hat{\lambda_{i}} - 
\lambda_{i}^{0})$ as:
\begin{eqnarray}
&&\langle O_{k}(\vec{V};\vec{\lambda}^{0}) \rangle_{\hat{\lambda}} =
\langle O_{k}(\vec{V};\vec{\lambda}^{0}) \rangle_{\vec{\lambda}^{0}} \nonumber \\
&&+
\sum_{i=1}^{\rho} \langle O_{i}(\vec{V};\vec{\lambda}^{0})\cdot
O_{k}(\vec{V};\vec{\lambda}^{0}) \rangle_{\vec{\lambda}^{0}} -
\langle O_{i}(\vec{V};\vec{\lambda}^{0}) \rangle_{\vec{\lambda}^{0}} \cdot
\langle O_{k}(\vec{V};\vec{\lambda}^{0}) \rangle_{\vec{\lambda}^{0}}] \cdot \hat{\Delta}_{i} \label{eq:25}
\end{eqnarray}
where the operation $\langle \Phi_{k}(\vec{V};\vec{\lambda}^{0}) \rangle_{\vec{\lambda}}$ denotes
the convolution:
\begin{equation}
\langle \Phi_{k}(\vec{V};\vec{\lambda}^{0}) \rangle_{\vec{\lambda}} = 
\int \Phi_{k}(\vec{V};\vec{\lambda}^{0}) \cdot g(\vec{V};\vec{\lambda}) d\vec{V}
\label{eq:25b}
\end{equation}
and the symbols with hats stand for the estimated quantities.
Thus, around the  initial value $\vec{\lambda}^{0}$, the right hand side 
of (\ref{eq:25}) can be easily
evaluated as a function of   the  values of the couplings (e.g. by using 
the four fermion M. C.
generators to produce events at $\vec{\lambda}^{0}$ couplings), 
whilst the left  hand side of (\ref{eq:25}) is estimated by using
the  $N$ kinematic vectors $\vec{V}_{n}$, $n= 1,\ldots ,N$  
measured by the experiment as:
\begin{equation}
\int O_{k}(\vec{V};\vec{\lambda}^{0}) \cdot g(\vec{V};\vec{\lambda}^{true}) d\vec{V} \simeq 
\frac{1}{N}\sum_{n=1}^{N}
 O_{k}(\vec{V}_{n};\vec{\lambda}^{0})
\label{eq:26}
\end{equation}
The  couplings values $\hat{\lambda} =\{ \hat{\lambda}_{1}, \ldots,
\hat{\lambda}_{\rho} \}$,
are then estimated by solving the linear system (of $\rho$ equations with 
$\rho$ unknowns) defined
by (\ref{eq:25}) and (\ref{eq:26}). Similarly the error in this estimation 
is evaluated in terms
of the the variance of the experimental measured quantities (\ref{eq:26}).

It must be emphasized that, due to the approximation  (\ref{eq:19}), this 
estimation is
consistent only when the true coupling values are close enough to the 
expansion point. 
In the asymptotic
limit, and when the above  condition holds, the estimation based on the mean 
value of the
optimal observables has the same\footnote{For a rather simple proof see at 
the Appendix A}
efficiency as the maximum likelihood. However, when dealing 
with
data sets of finite statistical size, special care must be taken to ensure 
that the approximation
(\ref{eq:19}) is valid for the region of the couplings values which 
correspond to mean values
of Optimal Observables ( via eq. \ref{eq:25}) lying within the uncertainty 
of the measured quantity
(\ref{eq:26}).  

In the case of small data sets, an increase in the estimating efficiency is 
expected by extending
the definition of the likelihood function  to count for the 
total number of the observed events.
The extended likelihood ($L^{ext}$) is defined as:
\begin{equation}
L^{ext}(\vec{V}_{1}, \ldots ,\vec{V}_{N};\vec{\lambda}) = 
\frac{\exp^{-\mu(\vec{\lambda})} 
\cdot \mu(\vec{\lambda})^{N}}{N!}\prod_{j=1}^{N}g(\vec{V}_{j},\vec{\lambda})
\label{eq:27}
\end{equation}
where $\mu(\vec{\lambda})$ is the expected number of  events for a 
luminosity ${\cal{L}}$ which is
dependent quadratically on the coupling values through equation (\ref{eq:5}).
The extended maximum likelihood equivalent estimators are formed (\cite{opob} 
and Appendix A) as the products
of the mean values of the Optimal Observables and the expected number of 
observed events. The estimation is then
performed by solving the non linear system of the following equations: 
\begin{eqnarray}
&&\sum_{n=1}^{N}
 O_{k}(\vec{V}_{n};\vec{\lambda}^{0}) = 
 \mu(\hat{\lambda})
\cdot [\langle O_{k}(\vec{V};\vec{\lambda}^{0}) \rangle_{\lambda^{0}} \nonumber \\
&&+
\sum_{i=1}^{\rho} [\langle O_{i}(\vec{V};\vec{\lambda}^{0})\cdot
O_{k}(\vec{V};\vec{\lambda}^{0}) \rangle_{\lambda^{0}} -
\langle O_{i}(\vec{V};\vec{\lambda}^{0}) \rangle_{\lambda^{0}} \cdot
\langle O_{k}(\vec{V};\vec{\lambda}^{0}) \rangle_{\lambda^{0}}] \cdot \hat{\Delta}_{i}
 \label{eq:29} \\
&& \mu(\vec{\lambda}) =
{\cal{L}} \cdot [
S_{0}+\sum_{k} S_{1}^{k}\cdot \vec{\lambda}_{k}+ \sum_{\mu k} 
S_{2}^{\mu k}\cdot \vec{\lambda}_{\mu}\vec{\lambda}_{k}]      \nonumber 
\end{eqnarray}
where, as in (\ref{eq:26}), the left hand side corresponds to the experimental 
measurement whilst the right hand side is evaluated as a function of the 
coupling values using the phenomenological models.

\section{The Realistic Case}

When the detector effects are sizable and cannot be ignored, 
the p.d.f. which contains the whole 
available information is written as:
\begin{equation}
P(\vec{\Omega},\vec{\lambda}) =
 \int \frac{c_{0}(\vec{V}) +\sum_{k} c^{k}_{1}(\vec{V}) \cdot\lambda _{k} + 
\sum_{\mu ,k} c_{2}^{\mu k}(\vec{V}) \cdot\lambda_{k} \cdot\lambda_{\mu} }
{\Theta_{0}+\sum_{k} \Theta_{1}^{k}\cdot\lambda_{k}+\sum_{\mu k} \Theta_{2}^{\mu k}\cdot\lambda_{\mu}
\cdot\lambda_{k}} \cdot \epsilon(\vec{V}) \cdot R(\vec{V},\vec{\Omega}) d\vec{V}
\label{eq:30}
\end{equation}
where
\begin{equation}
\Theta_{i}^{j} = \int c_{i}^{j}(\vec{V}) \cdot \epsilon(\vec{V})  d\vec{V}
\label{eq:31}
\end{equation}

\subsection{Optimal Variables Including Detector Effects}
Using the notation:
\begin{equation}
\tilde{c}_{i}^{j}(\vec{\Omega}) = \int c_{i}^{j}(\vec{V}) \cdot \epsilon(\vec{V}) \cdot R(\vec{V},\vec{\Omega}) 
d\vec{V}
\label{eq:31aa}
\end{equation}
the p.d.f. (\ref{eq:30}) is written as:
\begin{equation}
P(\vec{\Omega},\vec{\lambda}) =
  \frac{\tilde{c}_{0}(\vec{\Omega}) +\sum_{k} \tilde{c}^{k}_{1}(\vec{\Omega}) \cdot\lambda _{k} + 
\sum_{\mu ,k} \tilde{c}_{2}^{\mu k}(\vec{\Omega}) \cdot\lambda_{k} \cdot\lambda_{\mu} }
{\Theta_{0}+\sum_{k} \Theta_{1}^{k}\cdot\lambda_{k}+\sum_{\mu k} \Theta_{2}^{\mu k}\cdot\lambda_{\mu}
\cdot\lambda_{k}} 
\label{eq:30aa}
\end{equation}
which retain the functional form of (\ref{eq:7}). Consequently, 
in the case of a single TGC parameter model, the two directions of the phase 
space on which  the projected p.d.f.
retains the whole information will be:   
\begin{equation}
\zeta_{1}(\vec{\Omega}) = \frac{ \int c_{1}(\vec{V}) \cdot \epsilon(\vec{V}) \cdot R(\vec{V},\vec{\Omega}) d\vec{V}}
{ \int c_{0}(\vec{V}) \cdot \epsilon(\vec{V}) \cdot R(\vec{V},\vec{\Omega}) d\vec{V}}
\label{eq:32}
\end{equation}
\begin{equation}
\zeta_{2}(\vec{\Omega}) = \frac{ \int c_{2}(\vec{V}) \cdot \epsilon(\vec{V}) \cdot R(\vec{V},\vec{\Omega}) d\vec{V}}
{ \int c_{0}(\vec{V}) \cdot \epsilon(\vec{V}) \cdot R(\vec{V},\vec{\Omega}) d\vec{V}}
\label{eq:33}
\end{equation}

Equation (\ref{eq:32}) and (\ref{eq:33}) can be rewritten in  the general form:
\begin{equation}
\zeta_{1,2}(\vec{\Omega}) = \int Q_{1,2}(\vec{V}) \cdot D(\vec{V},\vec{\Omega}) d\vec{V} 
\label{eq:34}
\end{equation}
where the definitions of (\ref{eq:15q1}) and (\ref{eq:15q2}) have been used and 
$D(\vec{V},\vec{\Omega})$ stands for the following expression:
\begin{equation}
D(\vec{V},\vec{\Omega}) = \frac{ (c_{0}(\vec{V})/ \Theta_{0}) \cdot \epsilon(\vec{V}) \cdot R(\vec{V},\vec{\Omega}) }
{ \int (c_{0}(\vec{V})/ \Theta_{0}) \cdot \epsilon(\vec{V}) \cdot R(\vec{V},\vec{\Omega}) d\vec{V} }
\label{eq:35}
\end{equation}

The positive function $D(\vec{V},\vec{\Omega})$, is less or equal to one and 
normalized
to unity ($\int D(\vec{V},\vec{\Omega}) d\vec{V}$ = 1).
It expresses the conditional probability that:
{\em the kinematic
vectors $\vec{V}$  generated with the p.d.f. $c_{0}(\vec{V})/\Theta_{0}$ 
(i.e. with coupling equal to zero)
will be observed as $\vec{\Omega}$}. 
Consequently the variables $\zeta_{1,2}(\vec{\Omega})$ can been seen as the 
mean values of  $Q_{1,2}(\vec{V})$ with the condition that
the observed kinematic vectors be equal to $\vec{\Omega}$.
This interpretation of (\ref{eq:34}) suggests a pre-analysis stage, during which $\zeta_{1}$ and
$\zeta_{2}$ will be evaluated as functions of $\vec{\Omega}$ by  M.C. integration. However, the size
of the necessary sample of M. C. events increases exponentially with the 
dimensionality of the phase space and
thus, dealing with eight dimensional phase space, it makes this pre-analysis impractical.
The approach  followed in this work, was to project the resolution function $R(\vec{V},\vec{\Omega})$ on the
$c_{1}(\vec{\Omega})/c_{0}(\vec{\Omega})$, $c_{2}(\vec{\Omega})/c_{0}(\vec{\Omega})$ plane and  approximate the variables
$\zeta_{1,2}(\vec{\Omega})$ as:
\begin{equation}
\zeta_{1,2}(\vec{\Omega}) \simeq z_{1,2}(x_{1},x_{2}) = 
 \int Q_{1,2}(\vec{V}) \cdot \tilde{D}(\vec{V},x_{1},x_{2}) d\vec{V} \label{eq:36}
\end{equation}
\begin{equation}
 \tilde{D}(\vec{V},x_{1},x_{2}) =  \int D(\vec{V},\vec{\Omega}) \cdot \delta(x_{1}-c_{1}(\vec{\Omega})/c_{0}(\vec{\Omega}))
\cdot \delta(x_{2}-c_{2}(\vec{\Omega})/c_{0}(\vec{\Omega})) d\vec{\Omega}  \label{eq:37}
\end{equation}
where $\tilde{D}(\vec{V},x_{1},x_{2})$ is the 
the conditional probability that: {\em the kinematic
vectors $\vec{V}$, generated with the p.d.f.
$c_{0}(\vec{V})/\Theta_{0}$ 
and observed as $\vec{\Omega}$ correspond to values of 
$c_{1}(\vec{\Omega})/c_{0}(\vec{\Omega})$
and $c_{2}(\vec{\Omega})/c_{0}(\vec{\Omega})$ 
equal to $x_{1}$ and $x_{2}$ respectively}.
Figure \ref{o1resol} (and figure \ref{o2resol}) show the dependence  of $z_{1}(x_{1},x_{2})$
($z_{2}(x_{1},x_{2})$) on $c_{1}(\vec{\Omega})/c_{0}(\vec{\Omega})$ ($c_{2}(\vec{\Omega})/c_{0}(\vec{\Omega})$) in different
regions of $c_{2}(\vec{\Omega})/c_{0}(\vec{\Omega})$ ($c_{1}(\vec{\Omega})/c_{0}(\vec{\Omega})$) when
the $\alpha_{W\phi}$ TGC model is considered.
For the evaluation of these quantities  a  M.C. set of events $WW\rightarrow e\nu_e q\bar{q}$, 
produced with the EXCALIBUR \cite{exca} four fermion generator   with 
Standard Model (zero) couplings and  passed through the full
detector simulation program DELSIM \cite{delsim}, was used. 
Then the values of $z_{1,2}(x_{1},x_{2})$ which correspond to a region $\cal{B}$ of 
$[ c_{1}(\Omega )/c_{0}(\Omega ),c_{2}(\Omega )/c_{0}(\Omega )]$ plane were estimated according to (\ref{eq:36}) as
\begin{equation} 
z_{1,2}(x_{1},x_{2})\simeq \frac{1}{n_{\cal{B}}} \sum_{i=1}^{n_{\cal{B}}} Q_{1,2}(V_i)
\label{eq:36aa}
\end{equation}
where the reconstructed kinematic vectors $\Omega_i$ of the $n_{\cal{B}}$ M.C. events used in (\ref{eq:36aa})
were such that $c_{1}(\Omega_i)/c_{0}(\Omega_i),c_{2}(\Omega_i)/c_{0}(\Omega_i)$ belongs to $\cal{B}$.

The straight lines in these figures  indicate the region of equality
between the plotted quantities. The fact that the two quantities almost
coincide\footnote{Similar results have been obtained with the $\alpha_{W}$ and $\alpha_{B\phi}$ cross section
parameterization.}
suggests that the two directions of the phase space on which
the p.d.f. (\ref{eq:29}) retains the major part  of the information when projected, can be further approximated by:
\begin{equation}
\zeta_{1,2}(\vec{\Omega}) \simeq \frac{c_{1,2}(\vec{\Omega})}{c_{0}(\vec{\Omega})}
\label{eq:39}
\end{equation}   

\subsection{Modified Observables Including Detector Effects}

By expanding the probability distribution (\ref{eq:30}) around $\vec{\lambda}^{0}$ and
following the same arguments as in the previous section one finds that the mean values of the quantities
\begin{equation}
\omega_{k}(\vec{\Omega};\vec{\lambda}^{0})= \frac { \int y^{k}_{1}(\vec{V};\vec{\lambda}^{0}) \cdot \epsilon(\vec{V}) \cdot R(\vec{V},\vec{\Omega}) d\vec{V}}
{\int y_{0}(\vec{V};\vec{\lambda}^{0}) \cdot \epsilon(\vec{V}) \cdot R(\vec{V},\vec{\Omega}) d\vec{V}}
\label{eq:40}
\end{equation}
have the same estimating efficiency  as the unbinned likelihood for coupling 
values $\vec{\lambda}$ close to the expansion
point. These Optimal Observables $\omega_{k}(\vec{\Omega};\vec{\lambda}^{0})$ which include
the detector effects  can be expressed   as mean values under conditions, of the Optimal Observables 
$O_{k}(\vec{V};\vec{\lambda}^{0})$  defined  in the ideal case. This can been seen 
by rewriting (\ref{eq:40}) as:
\begin{equation}
\omega_{k}(\vec{\Omega};\vec{\lambda}^{0})= \int \frac{y_{1}^{k}(\vec{V};\vec{\lambda}^{0})}{y_{0}^{k}(\vec{V};\vec{\lambda}^{0})} \cdot B(\vec{V},\vec{\Omega}) d\vec{V}
\label{eq:41}
\end{equation}
with
\begin{equation}
B(\vec{V},\vec{\Omega}) = \int \frac{ (y_{0}^{k}(\vec{V};\vec{\lambda}^{0})/ \tilde{Z}_{0}(\vec{\lambda}^{0})) \cdot \epsilon(\vec{V}) \cdot R(\vec{V},\vec{\Omega})}
{ \int (y_{0}^{k}(\vec{V};\vec{\lambda}^{0})/ \tilde{Z}_{0}(\vec{\lambda}^{0})) \cdot \epsilon(\vec{V}) \cdot R(\vec{V},\vec{\Omega}) d\vec{V}} d\vec{V}
\label{eq:42}
\end{equation}
and
\begin{equation}
\tilde{Z}_{0}(\vec{\lambda}^{0})= \int y_{0}(\vec{V};\vec{\lambda}^{0}) \cdot \epsilon(\vec{V}) d\vec{V}
\end{equation}
where the conditional probability $B(\vec{V};\vec{\Omega})$  expresses the 
probability that: {\em the kinematic vectors $\vec{V}$ produced with p.d.f. 
$y_{0}^{k}(\vec{V};\vec{\lambda}^{0})/ \tilde{Z}_{0}(\vec{\lambda}^{0})$ 
(i.e. with coupling values equal to $\vec{\lambda}^{0}$) be observed as 
$\vec{\Omega}$}. The expected values of 
$\omega_{k}(\vec{\Omega};\vec{\lambda}^{0})$ for  coupling
values $\hat{\lambda}$ 
close to the expansion point $\vec{\lambda}^{0}$ are linearly dependent on $\hat{\lambda}$ as in (\ref{eq:25}).
Namely:
\begin{eqnarray}
\langle \omega_{k}(\vec{\Omega};\vec{\lambda}^{0}) \rangle_{\hat{\lambda}}=
\langle \omega_{k}(\vec{\Omega};\vec{\lambda}^{0}) \rangle_{\vec{\lambda}^{0}}&+&
\sum_{i=1}^{\rho} [\langle \omega_{k}(\vec{\Omega};\vec{\lambda}^{0}) \cdot 
\omega_{i}(\vec{\Omega};\vec{\lambda}^{0}) \rangle_{\vec{\lambda}_{0}} \nonumber \\ 
&-&\langle\omega_{k}(\vec{\Omega};\vec{\lambda}^{0})\rangle_{\vec{\lambda}_{0}}
 \cdot \langle \omega_{i}(\vec{\Omega};\vec{\lambda}^{0})\rangle_{\vec{\lambda}_{0}}] \cdot 
\hat{\Delta}_{i}
\label{eq:42a}
\end{eqnarray}
where the brackets stand for the operation: 
\begin{eqnarray}
\langle \Phi_{k}(\vec{\Omega};\vec{\lambda}^{0}) \rangle_{\vec{\lambda}}=
\int \Phi_{k}(\vec{\Omega};\vec{\lambda}^{0}) \cdot P(\vec{\Omega};\vec{\lambda}) d\vec{\Omega} 
 \label{eq:42c}
\end{eqnarray}

 The evaluation of the functional form of the Optimal Observable (\ref{eq:41}) in a preanalysis
stage faces  practical limitations, due to the   necessary number  of   M.C. 
events when dealing with
eight dimensional phase space. In this analysis, by projecting  the resolution function  on the
 $y_{1}^{k}(\vec{\Omega};\vec{\lambda}^{0})/y_{0}^{k}(\vec{\Omega};\vec{\lambda}^{0})$ axis, equation   
   (\ref{eq:41}) is approximated as:

\begin{equation}
\omega_{k}(\vec{\Omega};\vec{\lambda}^{0}) \simeq a_{k}(\phi;\vec{\lambda}^{0}) = 
 \int O_{k}(\vec{V};\vec{\lambda}^{0}) \cdot \tilde{B}(\vec{V},\vec{\phi}) d\vec{V} \label{eq:43}
\end{equation}
\begin{equation}
 \tilde{B}(\vec{V},\vec{\phi}) =  \int B(\vec{V},\vec{\Omega}) \cdot \prod_{k=1}^{\rho}
\delta(\phi_{k}-y^{k}_{1}(\vec{\Omega};\vec{\lambda}^{0})/y_{0}(\vec{\Omega};\vec{\lambda}^{0}))
 d\vec{\Omega}  \label{eq:44}
\end{equation}
\begin{equation}
\vec{\phi}=\{\phi_1,\ldots,\phi_{\rho} \} \label{eq:45}
\end{equation}
where $\tilde{B}(\vec{V},\phi;\vec{\lambda}^{0})$ is 
the conditional probability that: {\em the kinematic
vectors $\vec{V}$ generated with the p.d.f. 
$y_{0}(\vec{V};\vec{\lambda}^{0})/\tilde{Z}_{0}(\vec{\lambda}^{0})$ and 
observed as $\vec{\Omega}$  correspond to values of 
$y^{k}_{1}(\vec{\Omega};\vec{\lambda}^{0}/y_{0}(\vec{\Omega};\vec{\lambda}^{0})$ which are  
equal to $\phi^{k}$}. In figure \ref{y1awphi}, M.C. semileptonic events with 
an electron in the final state generated with the EXCALIBUR four fermion
generator and with Standard Model couplings, having passed through the full 
detector simulation,  are used to demonstrate
that for several initial values of the coupling   the  $\alpha_{W\phi}$ Optimal 
Observables can be further approximated\footnote{
This found to be a good approximation for the $\alpha_{W}$ and $\alpha_{B\phi}$ Observables as well.} as:
\begin{equation}
\omega_{k}(\vec{\Omega};\vec{\lambda}^{0}) 
\simeq \frac{y^{k}_{1}(\vec{\Omega};\vec{\lambda}^{0})}{y_{0}(\vec{\Omega};\vec{\lambda}^{0})} \label{eq:46}  
\end{equation}
Both approximations (\ref{eq:43}) and (\ref{eq:46}) respect the linear 
dependence of the mean values of the Optimal Observables on the true 
coupling values. However, the observables  (\ref{eq:46})\footnote{
In the following Modified Observables}  are not exactly optimal, in the sense
that  their mean values do not carry in principle the same  information  as the 
unbinned likelihood  function. The consistency and the unbiasedness, though, 
of these 
estimators are guaranteed by the inclusion of all the detector effects.

An estimation, based on the mean values of the Modified Observables,
inherits the limitations of the Optimal Observables described in the previous section. Namely
the linear dependence is valid only around the expansion point. In this paper we propose 
the evaluation of the mean of
the Modified Observable: 
\begin{eqnarray}
 \int (y^{k}_{1}(\vec{\Omega};\vec{\lambda}^{0}))/(y_{0} 
(\vec{\Omega};\vec{\lambda}^{0})) \cdot P(\vec{\Omega};\vec{\lambda}) d\vec{\Omega}
\nonumber
\end{eqnarray}
for any value $\vec{\lambda}$ by means of M.C. integration employing the reweighting technique \cite{rewght}.
This procedure is valid for any range of coupling values and consequently guarantees consistent 
estimation independently of the particular expansion point.
The above  arguments are demonstrated in
figure \ref{liner}  and  \ref{consi}.
A collection  of M.C. events, consisted from several sets produced at several 
couplings and passed through the full detector 
simulation\footnote{More details on these M.C. sets are 
given in Section 4.},  had been used to evaluate, by reweighting, 
the dependence (in the following the calibration curve)  of
the mean values of the Modified Observables for electronic final states on 
the $\alpha_{W\phi}$ coupling
for several ($\alpha_{W\phi}^{0}=$ -2,0 and 2) expansion points.
The points, in these figures, represent the calibration curves 
whilst the straight lines in figure  \ref{liner} correspond to 
the linear equation (\ref{eq:42a}).
It is, though, obvious that the consistency of an  estimation based on the 
linear relation (\ref{eq:42a}), as proposed in 
\cite{opob}, depends strongly on the expansion point. 
However, when the calibration curve is used, the estimation is consistent
independently of the expansion point, as is shown in figure  \ref{consi} where a set of a
M.C. electronic events have been used as a data sample.
These events  were generated with the EXCALIBUR four fermion generator with 
Standard Model couplings
and  passed through a  full detector simulation. The  mean values of the 
Modified Observables evaluated using the data sample as follows:
\begin{equation}
\langle \omega_{k}(\vec{\Omega};\alpha_{W\phi}^{0}) \rangle \simeq
\frac{1}{N} \sum_{n=1}^{N} \frac{y_{1}(\Omega_{n};\alpha_{W\phi}^{0})}{y_{0}(\Omega_{n};\alpha_{W\phi}^{0})}
\label{eq:48}
\end{equation}
are shown in figures \ref{consi}a),
\ref{consi}b) and \ref{consi}c) as
the central horizontal lines 
whilst the band around the central line corresponds to the statistical error 
of (\ref{eq:48}).
Although the calibration curves coincide with the measured averages at the true 
coupling value independently of the expansion point,
the error in the coupling estimation  varies,
\footnote {In this example, for $N = 1000$, the errors were 0.1,0.07 and 0.1 for $\alpha_{W\phi}^{0}$ -2, 0 and 
+2 respectively.}
reaching its minimum at the expansion point which equals the true coupling value. 

The extended Modified Observable estimators which take into account the expected event multiplicity
 are built  in the same way as in the ideal case, i.e.
as the product: 
\begin{eqnarray}
\mu(\vec{\lambda}) \cdot \int \omega_{k}(\vec{\Omega};\vec{\lambda}^{0}) \cdot P(\vec{\Omega};\vec{\lambda})
d\vec{\Omega} 
\nonumber
\end{eqnarray}

The  calibration functions 
\begin{equation}
{\cal{F}}_{k}(\vec{\lambda};\vec{\lambda}^{0}) = \mu(\vec{\lambda}) \cdot \int \omega_{k}(\vec{\Omega};\vec{\lambda}^{0}) 
\cdot P(\vec{\Omega};\vec{\lambda})d\vec{\Omega}
\label{eq:49}
\end{equation}
are evaluated as before by reweighted M.C. integration and 
the coupling estimates are the solutions of 
the following system of $\rho$ equations
with $\rho$ unknowns:
\begin{equation}
\sum_{n=1}^{N} \frac{y_{1}^{k}(\vec{\Omega}_{n};\vec{\lambda}^{0})}{y_{0}
(\vec{\Omega}_{n};\vec{\lambda}^{0})} =
{\cal{F}}_{k}( \hat{\lambda};\vec{\lambda}^{0})
\label{eq:50}
\end{equation}
Since the  estimation efficiency
becomes optimal only when $\hat{\lambda} \simeq \vec{\lambda}^{0}$, the
estimation being consistent for each $\vec{\lambda}^{0}$, 
an iterative procedure can be followed which 
will converge to the above optimality condition. 
To demonstrate the convergence properties of this
algorithm, three M.C. samples of 1000 $WW\rightarrow e \nu_e q \bar{q}$  events each, passed 
through detector simulation
and generated with the  EXCALIBUR four fermion generator with
$\alpha_{W\phi}$ coupling values equal to -2,0 and +2 respectively, were
fitted  by solving equation (\ref{eq:50}) for several values of the expansion point ($\alpha_{W\phi}^{0}$).
After every fit, the difference between the estimated value of the coupling 
and the expansion value  $\alpha_{W\phi}^{0}$
defines the step of the iterative procedure. Convergence is achieved at the expansion point where
the  step equals to zero.
As it is shown in figure  \ref{iter}, convergence is achieved after few only (two) iterations
almost independently of the initial values.

In the general case, where the data sample consists of a collection of the 
three semileptonic channels (muonic, electronic and tau),  the p.d.f.
is the following weighted sum of the individual distributions:
\begin{eqnarray}
P(\vec{\Omega} ;\vec{\lambda} ) &=& \sum_{f=1}^{3}w_{f}(\lambda)
\cdot\int g_{f}(\vec{V};\vec{\lambda})\cdot R_{f}(\vec{V},\vec{\Omega})d\vec{V} \nonumber \\
&=& \sum_{f=1}^{3} w_{f}(\lambda)\cdot P_{f}(\vec{\Omega} ;\vec{\lambda} ) \nonumber \\
P_{f}(\vec{\Omega} ;\vec{\lambda} ) &=& \int g_{f}(\vec{V};\vec{\lambda})\cdot R_{f}(\vec{V},\vec{\Omega})d\vec{V}
\nonumber \\
w_{f}(\lambda) &=& \frac{\sigma_{obs}^{f}(\vec{\lambda})}{\sum_{f=1}^{3}\sigma_{obs}^{f}(\vec{\lambda})}
\label{eq:2dd}
\end{eqnarray}
where the subscript $f$ stands for the flavor of the final state lepton. 
Then the extended Modified Observable
estimators are also defined as  weighted sums of the form:
\begin{equation}
M(\vec{\lambda}) \cdot
\sum_{f=1}^{3}w_{f}(\vec{\lambda}) \cdot \int \omega_{k,f}(\vec{\Omega};\vec{\lambda}^{0}) 
\cdot P_{f}(\vec{\Omega};\vec{\lambda})d\vec{\Omega} = \sum_{f=1}^{3}\mu_{f}(\vec{\lambda})
 \cdot \int \omega_{k,f}(\vec{\Omega};
\cdot P_{f}(\vec{\Omega};\vec{\lambda})d\vec{\Omega}
\label{eq:2ee}
\end{equation}
where $M(\vec{\lambda})$ stands for the expected number of the  events in total, 
whilst $\mu_{f}(\vec{\lambda})$ denotes the expected number of  events in the semileptonic channel $f$.
In practice one has  to evaluate three sets of calibration curves ${\cal{F}}_{k,f}(\vec{\lambda},\vec{\lambda}^{0})$,
one for each channel,
and to use the $N(=\sum_{f=1}^{3} N_{f})$ selected events to solve the following system of equations:

\begin{equation}
\sum_{{f}=1}^{3}\sum_{n=1}^{N_{f}} \frac{y_{1,f}^{k}(\vec{\Omega}_{n};\vec{\lambda}^{0})}{y_{0,f}
(\vec{\Omega}_{n};\vec{\lambda}^{0})} =
\sum_{f=1}^{3}{\cal{F}}_{k,f}( \hat{\lambda};\vec{\lambda}^{0})
\label{eq:50aa}
\end{equation}

\section{Numerical Results}

In the previous section we proposed two strategies which include the detector effects in efficient
estimators. In principle 
their asymptotic efficiency is less than the unbinned eight-dimensional 
likelihood estimator due to the projections
(\ref{eq:36}) and (\ref{eq:43}). There is  not a simple way to quantify the loss in
 information  due to the fact
that an unbinned  maximum likelihood 
in eight dimensions estimator which includes the detector  effects is 
practically impossible to build. Therefore, it makes more sense to discuss in  
detail the use and properties of these estimators in extracting the TGC's 
especially
when one deals with small data samples such as the available data samples at 
172 $GeV$ 
centre of mass energy at LEPII. The examples given bellow concern one parameter fits of different sensitivities,
where the estimation of the $\alpha_{W\phi}$, $\alpha_{W}$ and $\alpha_{B\phi}$ couplings
have been chosen for this purpose.

In all the following demonstrations the integrations performed by M.C. techniques
made use of the reweighting procedures  to express accurately  integrals as functions of the TGC's.
Several M.C. samples, generated either with the PYTHIA \cite{pythia} (including only the resonant
graphs, and ISR) or the EXCALIBUR (including the full set of four fermion graphs and ISR) generators 
at different coupling values and having passed through the detector simulation program
DELSIM,
were reweighted to correspond to the full set of four fermion diagrams with Coulomb corrections.
These samples have been combined  as in \cite{rewght}, 
in order to increase the statistical
accuracy of the reweighted M.C. integrations.
In parallel other M.C. set of events, produced with the EXCALIBUR four fermion generator at  
certain coupling values, played the role of data sets after 
being passed through
DELSIM. The selection criteria,  applied to all the M.C. events, 
were those described in Section 2.
The event multiplicity of each of the M.C. data samples was chosen 
according to Poissonian distributions with mean values
corresponding to the expected number of events after reconstruction with integrated luminosity $\sim ~10~ pb^{-1}$ (a typical integrated luminosity received by the LEP experiments at $\sqrt{s}~=~172~ GeV$).
 As an example, the data sets which were produced with Standard Model couplings consist of
$n_{\mu}+n_{e}+n_{\tau}$ events, where the subscript denotes the lepton flavor in the final state and the 
$n_{\mu},n_{e},n_{\tau}$  multiplicities varied from set to set 
following Poissonian distributions with means 15.35, 12.5 and 5.64
respectively.

\subsection{Binned Likelihood Fits}
 The central assumption in this paper is that the resolution and the 
selection efficiency 
can not be expressed easily as  functions of eight kinematic variables. 
Consequently the projected p.d.f. in
the $\zeta_{1}(\vec{\Omega})$,  $\zeta_{2}(\vec{\Omega})$ plane 
(\ref{eq:39}) cannot be expressed analytically. Its numerical evaluation is,
though, possible using M.C. events. There are several techniques 
\cite{bishop} of functional
interpolation but this work followed the simplest, namely, the determination of 
the projected distribution in bins of
$\zeta_{1}$ and $\zeta_{2}$. Then the extended likelihood function, when N events are observed,  distributed as 
$n_{k}$ ($k=1,\ldots 3\cdot\beta$) events (including background contribution) 
in the $k^{th}$ $\{ \zeta_{1},\zeta_{2}\}$ bin, is written as:     

\begin{eqnarray}
L^{ext} = \prod_{k=1}^{3 \cdot \beta} 
\frac{1.}{2 \cdot \pi \cdot \sigma_{k}(\lambda) \sigma_{bk}}
\int \int \frac{(x_{k}+y_{k})^{n_{k}}}{n_{k}!}
e^{-(x_{k}+y_{k})} 
e^{- \frac{(x_{k}-\mu_{k}(\lambda))^{2}}{{2\cdot \sigma_{k}}^{2}(\lambda)}} 
e^{- \frac{(y_{k}-b_{k})^{2}}{{2\cdot \sigma_{bk}}^{2}}}dx_{k} \cdot dy_{k}
\label{eq:49bbb}
\end{eqnarray}
Each of the $\beta$ consecutive terms in this product belongs  to one  of the three semileptonic
final states, whilst       
$\mu_{k}(\lambda)$ ($b_{k}$) and $\sigma_{k}(\lambda)$ ($\sigma_{bk}$) are 
the expected number
of signal (background) events  in the $k^{th}$ bin and the gaussian error  
respectively.
The evaluation of $\mu_{k}(\lambda)$ and its error as a function of the 
couplings is done by reweighted M.C. integration 
where the detector effects and the contamination of each of the final state channels
from each other have been  taken into account. A technical detail,
worth mentioning, is the fact
that the  calculation of the coordinates $\zeta_{1,2}$  from the observed vector $\Omega$ is solely based on 
the Matrix Elements because other factors, such
as the phase space and that expressing the initial state radiation,
 cancel out in (\ref{eq:39}).

To demonstrate the statistical properties of this technique, 210 
M.C. data samples  at Standard Model Couplings and without background contribution were fitted. The produced distributions of 
the estimated 
$\alpha_{W \phi}, \alpha_{W},  \alpha_{B\phi}$  in one-TGC model fits, are shown in figures 
\ref{leisos_awphi}a,
\ref{leisos_aw}a and \ref{leisos_bphi}a. These distributions are found to be in excellent 
agreement with Gaussians of
means ($0.01 \pm 0.02$, $0.007 \pm 0.04$ and $0.07 \pm 0.1$)
consistent with the true coupling values and sigmas equal to  $0.34 \pm 0.02$,$0.6 \pm 0.03$,$1.25 \pm 0.05$ 
respectively.
 Furthermore the pull distributions\footnote{Which is the distribution of the 
deviation of
each individual estimation from the true coupling normalized to the estimated 
error.}, shown in   \ref{leisos_awphi}b,
\ref{leisos_aw}b and \ref{leisos_bphi}b are found to be normal 
with sigmas ($0.94 \pm 0.05$, $1.02 \pm 0.06$ and $0.97 \pm 0.05$) 
consistent with unity, which indicates that the errors of the estimations are 
correctly evaluated. This is also
supported by the very good agreement of the sigma of the distribution of the  estimations
and the mean of the  distribution of the estimated error in each individual fit ($0.34 \pm 0.04$, 
$0.58 \pm 0.05$ and $1.28 \pm 0.02$). 
Similar tests performed with 20  sets  produced with $\alpha_{W \phi}$ 
values at -2, and 2 demonstrated the same properties.
\footnote{ At $\alpha_{W \phi}=-2$ the average  of the estimations are  
$-1.99 \pm 0.06$ whilst the root mean squared pull is $0.96 \pm 0.16$. Similar
tests at $\alpha_{W \phi}=+2$ gave as mean of the estimations $1.98 \pm 0.06$ 
with a root mean squared of the pulls $1.007 \pm 0.160$.}.

\subsection{Modified Observable Fits}

An estimation of the couplings based on the extended Modified
Observable technique is expected to be at least equally efficient as the binned extended likelihood 
estimation of the previous section. Furthermore, the Modified Observable fits
do not split the events into bins and in principle they suffer from less systematic error due to the
M.C. statistics.
However, as has been emphasized in the previous sections, the optimal efficiency of this
technique is achieved when the uncertainty on the measurement of the mean 
of the Modified
Observables lies within the linear part of the calibration curve around the expansion point.

The  calibration curve (${\cal{F}}(\lambda;\lambda^{0})$) and its 
statistical error(${\cal{E}}(\lambda;\lambda^{0})$) at the $\lambda^{0}$ expansion point are evaluated
by reweighted M.C. integration as  functions of the fitted coupling. In the general case, 
taking also into account the statistical errors due to the finite M.C. statistics, the coupling value is 
extracted by maximizing the following likelihood function:   
\begin{eqnarray}
L(\lambda;\lambda_{0}) =
\frac{1} {\sqrt{2 \cdot \pi \cdot 
(\sigma_{d}^{2}(\lambda_{0}) +\sigma_{b}^{2}(\lambda_{0})+
{\cal{E}}^{2}(\lambda;\lambda_{0}))}}
\cdot e^{A(\lambda;\lambda_{0})} \label{eq:60a}
\end{eqnarray}
where
\begin{eqnarray}
A(\lambda;\lambda_{0}) = {- \frac{ ( [N \cdot \langle \omega(\Omega;\lambda_{0}) 
\rangle _{data} - N_{b} \cdot 
\langle \omega(\Omega;\lambda_{0}) \rangle _{back} ] - 
{\cal{F}}(\lambda;\lambda_{0}) )^{2}}
{2 \cdot (\sigma_{d}^{2}(\lambda_{0})+\sigma_{b}^{2}(\lambda_{0}) +{\cal{E}}^{2}(\lambda;\lambda_{0}))} }
\label{eq:60}                     
\end{eqnarray}
and  N is the number of selected events with measured vectors $\{\vec{\Omega}_{1}, \ldots \vec{\Omega}_{N}\}$,
$N_{b}$ is the number of background events expected in
the data sample and $\sigma_{d}(\lambda_{0})$ is the measurement error on the quantity
$N \cdot \langle \omega(\Omega;\lambda^{0})\rangle _{data}$ which is evaluated from the data as:

\begin{equation}
N\cdot \langle \omega(\Omega;\lambda_{0}) \rangle _{data} \simeq  \sum_{f=1}^{3}\sum_{n=1}^{N_{f}} \cdot 
\frac {y_{1,f}(\Omega_{n};\lambda_{0})} {y_{0,f}(\Omega_{n};\lambda_{0})}
\label{eq:61}
\end{equation}
The background term and its error, $\sigma_{b}(\lambda_{0})$ are calculated by using a set of M background
M.C. events as:
\begin{equation}
N_{b} \cdot\langle \omega(\Omega;\lambda_{0}) \rangle _{back} \simeq \frac{N_b}{M} \sum_{i=1}^{M} \cdot 
\frac {y_{1}(\Omega_{i};\lambda_{0})} {y_{0}(\Omega_{i};\lambda_{0})}
\end{equation}
As discussed in the previous sections, an iterative procedure has to be followed until the value of
the coupling which maximizes (\ref{eq:60}) coincides with the expansion point.

As in the case of the binned likelihood estimation, the 210 M.C. sets produced with
Standard Model couplings and without background contamination were fitted to determine
the $\alpha_{W\phi}$, $\alpha_{W}$ and $\alpha_{B\phi}$ couplings. The results are shown in 
Figures \ref{sample_awphi}, \ref{sample_aw} and \ref{sample_bphi} where
Gaussian fits have been performed to the estimation, pull and estimation error distribution.
The technique exhibits the desired properties at least for the case of the $\alpha_{W\phi}$ and
$\alpha_{W}$ fits, that is the estimation distributions are found consistent 
with Gaussians 
with means at $\langle \hat{\alpha}_{W\phi} \rangle =-0.01\pm0.02$ and
$\langle \hat{\alpha}_{W} \rangle =0.05\pm0.05$ (which indicates the unbiasedness of the estimation) and
sigmas $\sigma_{\alpha_{W\phi}}=0.33\pm0.02$ and $\sigma_{\alpha_{W}}=0.56\pm0.03$ which are 
in excellent agreement
with the mean of the distribution of the estimated errors in each individual fit, $0.33\pm0.04$
and $0.58\pm0.01$ respectively. The correct estimation of the errors in each fit is also demonstrated by
the fact that the pull distributions are normal with sigmas consistent with unity ($1.05\pm0.07$ for
the $\alpha_{W\phi}$ and $0.95\pm 0.06$ for the $\alpha_{W}$ fits). The same properties  have been 
found in estimating $\alpha_{W\phi}$ by using the 20 sets of events produced at -2 and +2 coupling values
\footnote{ The mean values of the estimations are $-1.99 \pm0.02$ and $2.024\pm0.022$
whilst the root mean squared pulls are $0.9\pm0.15$ and $0.88\pm0.16$ for -2 and +2
true coupling values respectively.}. 
However, the fits to extract the $\alpha_{B\phi}$ coupling make apparent that 
there is  a limitation to this
technique. Although the distribution of the estimations has a Gaussian shape with mean consistent
with the true coupling ($0.08\pm0.10$) and a sigma equal to $1.25\pm0.06$,
the pull distribution deviates from a Gaussian shape due to an 
excess of values at pull values around zero. Furthermore the distribution of the estimated errors
consists of a gaussian peak centered around 1.3 and of a broad shoulder at higher values
which indicates that for a fraction of the fits the error was overestimated. 
This is a direct consequence of a large statistical error in evaluating the 
experimental quantity (\ref{eq:61}) compared  to the region in which the approximation
(\ref{eq:19}) is valid \footnote{ This can be seen by enlarging the error bands in figure \ref{consi}}.

\subsection{Comparison of Fitting Techniques}
In order to quantify the loss in precision due to the approximations  
(\ref{eq:36}) and (\ref{eq:43}), the 210 M.C. samples of events used to investigate the properties
of the fitting techniques proposed in this paper were used also in unbinned extended likelihood fits by using
either the true kinematic vectors of the events (in the following perfect extended likelihood fits) or the
reconstructed kinematic vectors after the event has been treated in a 6c constrained fit (in the
following 6c unbinned extended likelihood fits). The results of these fits 
as well as the results of the binned likelihood and the extended Modified Observables fits described
in the previous sections are summarized in Table
\ref{tab1}. Although the available statistical accuracy provided by this M.C. experimentation
with only 210 samples is not enough for strong statements, it can be seen
that the 6c unbinned extended likelihood
fits, suffer
from biases in estimating 
the $\alpha_{w\phi}$ and $\alpha_{w}$ couplings. These biases are eliminated by application of our proposed
techniques.

Since the p.d.f (\ref{eq:3})
carries less information than the p.d.f (\ref{eq:2}), it is to be expected that the
precisions obtained by the binned 2-dimensional likelihood 
and the extended Modified Observable methods should be somewhat
poorer than those from the perfect EML method. The statistics used for
the results of Table \ref{tab1} are insufficient to demonstrate this conclusively;
however, it is clear that the loss of information is very moderate. 

\section{Conclusions}

In this paper it has been shown that the reduction of the necessary kinematic
variables in the estimation of TGCs is possible
without loosing accuracy. This projection makes it possible
to include the detector effects in a binned extended likelihood estimation by employing
the reweighting \cite{rewght} M.C. integration technique. Although this projection is useful only
in one TGC parameter fits, an extension of the Optimal Observable technique 
\cite{opob} proposed in this work
can further reduce the necessary kinematic parameters to as many Modified 
Observables
as the number of the TGC's which are fitted simultaneously from the data, 
while continuing to include the detector effects
in the estimators. It is also shown that, by using the reweighting technique, an iterative procedure can be defined
with optimal convergence properties.

Both the proposed fitting strategies have been demonstrated to be unbiased estimators when  small statistical samples, of the same
size as the data sets
available from the 172 $GeV$ LEPII run, were fitted. It has been also 
shown that they consistently
evaluate  the estimation error. However a limitation of the Modified Observable fits
became apparent in evaluating the error in the fitted value of the $\alpha_{B\phi}$ coupling.
Such an overestimation is expected in
fits where the data are not sensitive to the extracted parameter and will 
disappear as the
available data sets increase in size.

A comparison of the proposed techniques with the unbinned  extended 
likelihood in perfect conditions shows
that the estimation accuracy achieved in this analysis is very close to the maximum possible.

\appendix
\section{Appendix A}
 For $\hat{\lambda} = \vec{\lambda}^{true}$ equation (\ref{eq:11b}) becomes identity. Furthermore by using
(\ref{eq:8}) and approximating $g(\vec{V};\vec{\lambda})$  as in (\ref{eq:19})  the definition of the maximum
likelihood estimator relates the mean values of the Optimum observables ($y_{1}^{i}(\vec{V};\vec{\lambda}^{0})/
y_{0}(\vec{V};\vec{\lambda}^{0})$ as:
\begin{eqnarray}
&&\int \{\frac{y_{1}^{i}(\vec{V};\vec{\lambda}^{0})/y_{0}(\vec{V};\vec{\lambda}^{0}) - Z^{i}_{1}(\vec{\lambda}^{0})
/Z_{0}(\vec{\lambda}^{0})}
{1+\sum_{k=1}^{\rho} [y_{1}^{k}(\vec{V};\vec{\lambda}^{0})/y_{0}(\vec{V};\vec{\lambda}^{0}) - 
Z^{k}_{1}(\vec{\lambda}^{0})/Z_{0}(\vec{\lambda}^{0})] \cdot
\hat{\Delta}_{k}} \pm \frac{y_{1}^{i}(\vec{V};\vec{\lambda}^{0})}
{y_{0}(\vec{V};\vec{\lambda}^{0})} \} \cdot \nonumber \\ 
&&\{\frac{y_{0}(\vec{V};\vec{\lambda}^{0})}{Z_{0}(\vec{\lambda}^{0})} \cdot [ 
1+\sum_{k=1}^{\rho} [y_{1}^{k}(\vec{V};\vec{\lambda}^{0})/y_{0}(\vec{V};\vec{\lambda}^{0}) - 
Z^{k}_{1}(\vec{\lambda}^{0})/Z_{0}(\vec{\lambda}^{0})] \cdot
\hat{\Delta}_{k} ]\} d \vec{V}  = 0 \label{eq:a1}
\end{eqnarray}

where $\hat{\Delta}_{k} = \hat{\lambda}_{k}-\lambda^{0}_{k}$
and the term 
$y_{1}^{i}(\vec{V};\vec{\lambda}^{0})/
y_{0}(\vec{V};\vec{\lambda}^{0})$ has been added and subtracted in the integrand.
 Then, by expressing the ratio of the integrated quantities as:
\begin{equation}
 \frac {Z^{i}_{1}(\vec{\lambda}^{0})}{Z_{0}(\vec{\lambda}^{0})} = \int 
\frac{y_{1}^{i}(\vec{V};\vec{\lambda}^{0})}{y_{0}(\vec{V};\vec{\lambda}^{0})} \cdot 
\frac{y_{0}(\vec{V};\vec{\lambda}^{0})}{Z_{0}(\vec{\lambda}^{0})} d\vec{V} \label{eq:a2}
\end{equation}
and using (\ref{eq:24a})  the identity (\ref{eq:a1}) becomes :

\begin{eqnarray}
&&\int O_{i}(\vec{V};\vec{\lambda}^{0}) \cdot g(\vec{V};\hat{\lambda}) d\vec{V} = \nonumber \\
&&\int O_{i}(\vec{V};\vec{\lambda}^{0}) \cdot g(\vec{V};\vec{\lambda}^{0}) 
d\vec{V} 
 +
\sum_{k=1}^{\rho} [\int O_{i}(\vec{V};\vec{\lambda}^{0}) \cdot O_{k}(\vec{V};\vec{\lambda}^{0}) 
\cdot g(\vec{V};\vec{\lambda}^{0})d\vec{V}  \nonumber \\
&&-(\int O_{i}(\vec{V};\vec{\lambda}^{0}) \cdot 
g(\vec{V};\vec{\lambda}^{0})d\vec{V}) \cdot
(\int O_{k}(\vec{V};\vec{\lambda}^{0}) \cdot 
g(\vec{V};\vec{\lambda}^{0})d\vec{V})
]\cdot \hat{\Delta_{i}}
 \label{eq:a3}
\end{eqnarray}
which is exactly how the coupling estimators based on the mean of optimal observables have been defined in
(\ref{eq:25}). Thus in the neighborhood of $\vec{\lambda}^{0}$ ( where the approximation (\ref{eq:19}) is valid)
the maximum likelihood estimation is identical to the Optimal Observable estimations and obviously the latter inherits
all the properties of the former.  

If instead of the likelihood function its extension (\ref{eq:27}) was used then the extended maximum likelihood estimators
will be defined for $N$ observed events as:
\begin{eqnarray}
&&\sum_{n=0} [\frac{e^{-\mu(\vec{\lambda})}\cdot \mu^{N}}{N!}]_{\vec{\lambda}=\vec{\lambda}^{true}} \cdot
\{ [\frac{\partial}{\partial{\lambda_{i}}}(\frac{e^{-\mu(\vec{\lambda})}\cdot \mu^{N}}{N!})]_{\vec{\lambda}=\hat{\lambda}}
 \nonumber \\
&&+\int [\frac{\partial}{\partial{\lambda_{i}}} \sum_{k=1}^{N} \log{g(\vec{V}_{k};\vec{\lambda})}]_{\vec{\lambda}=\hat{\lambda}} 
\cdot [ \prod_{j=1}g(\vec{V}_{k};\vec{\lambda})]
_{\vec{\lambda}=\vec{\lambda}^{true}}d\vec{V}_{1} \cdots d\vec{V}_{N} 
 \}
 = 0\label{eq:a4}
\end{eqnarray}
The above condition obviously for $\hat{\lambda} = \vec{\lambda}^{true}$ is an identity which
can be reduced, after some algebra, to:
\begin{equation}
\mu(\hat{\lambda}) \cdot \int  
[\frac{\partial}{\partial{\lambda_{i}}} \log{g(\vec{V};\vec{\lambda})}]
\cdot g(\vec{V};\vec{\lambda})]_{\vec{\lambda}=\hat{\lambda}} d\vec{V} =0
\label{eq:a5}
\end{equation}
and when the p.d.f is linearly approximated as in (\ref{eq:19}) the same expression as (\ref{eq:a3}) is achieved
having  replaced the terms $O_{i}(\vec{V};\vec{\lambda}^{0})$ with the products
$\mu(\hat{\lambda})\cdot O_{i}(\vec{V};\vec{\lambda}^{0})$.


\begin{table}[h]
\begin{center}
\begin{tabular}{|c|c|c|c||}  
\hline
      & $\alpha_{w\phi}$ & $\alpha_{w}$ & $\alpha_{B\phi}$ \\
\hline
Perfect Extended Likelihood &      &       &        \\
\hline
mean of estimations & 0.006 $\pm$ 0.020 & 0.01 $\pm$ 0.03 & 0.11 $\pm$ 0.08 \\
estimation accuracy & 0.30 $\pm$ 0.02       & 0.55 $\pm$ 0.03      &  1.10 $\pm$ 0.05     \\
\hline
6c Unbinned Extended Likelihood &           &        &          \\
\hline
mean of estimations & -0.07 $\pm$ 0.02 & 0.14 $\pm$ 0.04 & 0.001 $\pm$ 0.080\\
estimation accuracy & 0.35$\pm$ 0.02   & 0.65 $\pm$ 0.03         &  1.14 $\pm$ 0.06           \\
pull sigma        & 1.10 $\pm$ 0.05 & 1.35 $\pm$ 0.06 & 1.03 $\pm$ 0.05\\
\hline
Binned 2 dim. Extended Likelihood  &                  &          &              \\
\hline
mean of estimations & 0.01 $\pm$ 0.02 & 0.007 $\pm$ 0.040 & 0.07 $\pm$ 0.10\\
estimation accuracy & 0.34 $\pm$ 0.02 & 0.60 $\pm$ 0.04  & 1.25$\pm$ 0.08\\
pull sigma        & 0.94 $\pm$ 0.05 & 1.02 $\pm$ 0.06 & 0.97 $\pm$ 0.05\\
\hline
Extended Modified  Observable   &                  &          &              \\
\hline
mean of estimations & -0.01 $\pm$ 0.02 & 0.05 $\pm$ 0.05 & 0.08 $\pm$ 0.1\\
estimation accuracy & 0.33 $\pm$ 0.02  & 0.56 $\pm$ 0.03 & 1.25$\pm$ 0.06  \\
pull sigma        & 1.05 $\pm$ 0.07 & 0.95 $\pm$ 0.05 & $0.90^{*}$ $\pm$ 0.05 \\
\hline
\end{tabular} 
\end{center} 
\caption{Comparison of the statistical properties of
the techniques proposed in this paper with the unbinned extended
likelihood estimations.(*This value represents
the root mean squared of the pulls)}
{\label{tab1}} 
\end{table}

\clearpage    
\begin{figure}[o1resol]
\centerline{\epsfig{file=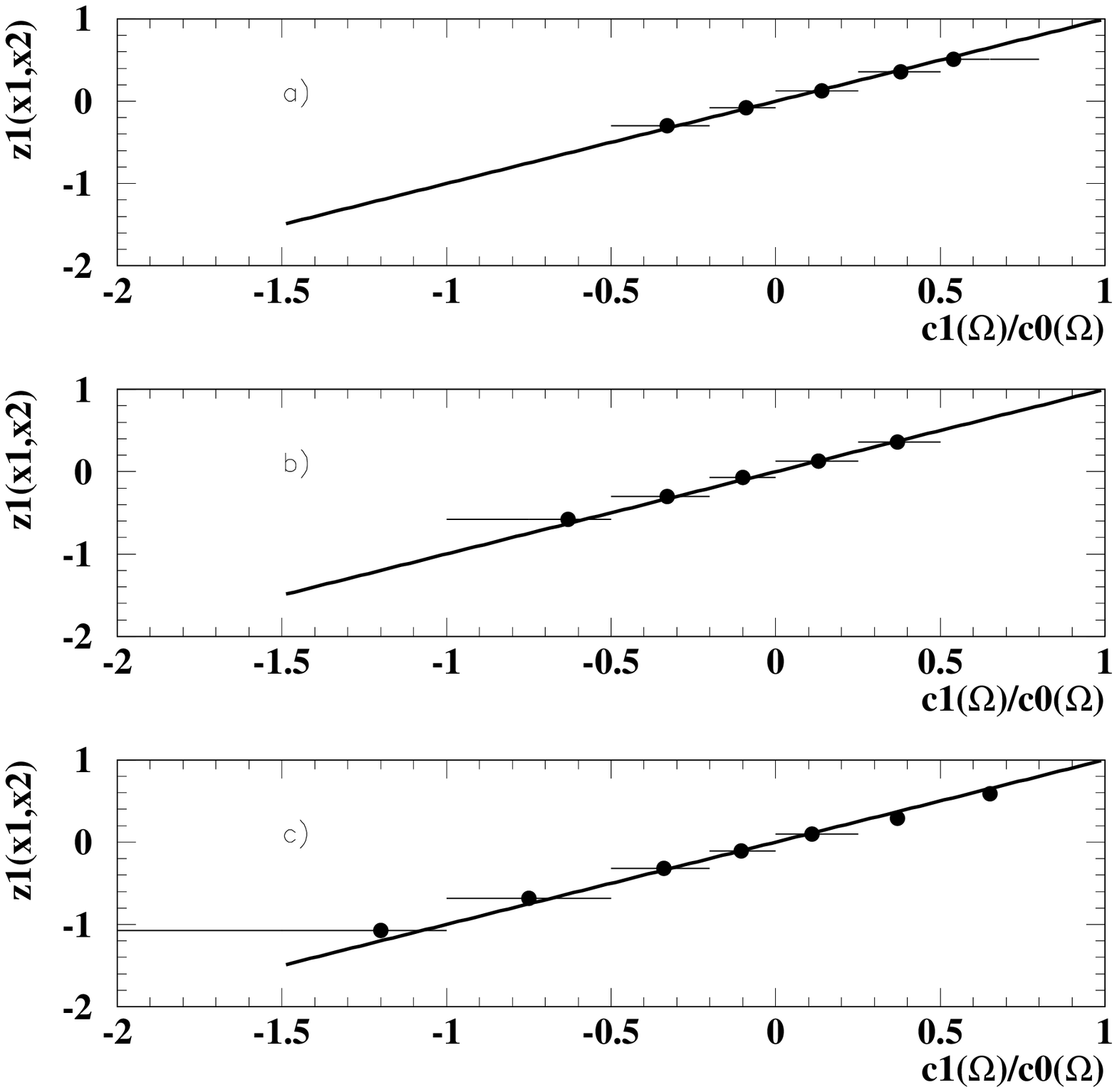,height=18cm}}
\caption{ 
The dependence  of $z_{1}(x_{1},x_{2})$
on the $c_{1}(\vec{\Omega})/c_{0}(\vec{\Omega})$  in three different
regions (a) 0.00-0.12, b) 0.12-0.25 and c) 0.25-1.00) of $c_{2}(\vec{\Omega})/c_{0}(\vec{\Omega})$  when
the $\alpha_{W\phi}$ TGC model is consider for electronic final states. 
The horizontal error-bars
correspond to the size of the bin, whilst the solid line denotes 
the region of equality between the two plotted quantities.}
{
\label{o1resol}}
\end{figure}
\begin{figure}[o2resol]
\centerline{\epsfig{file=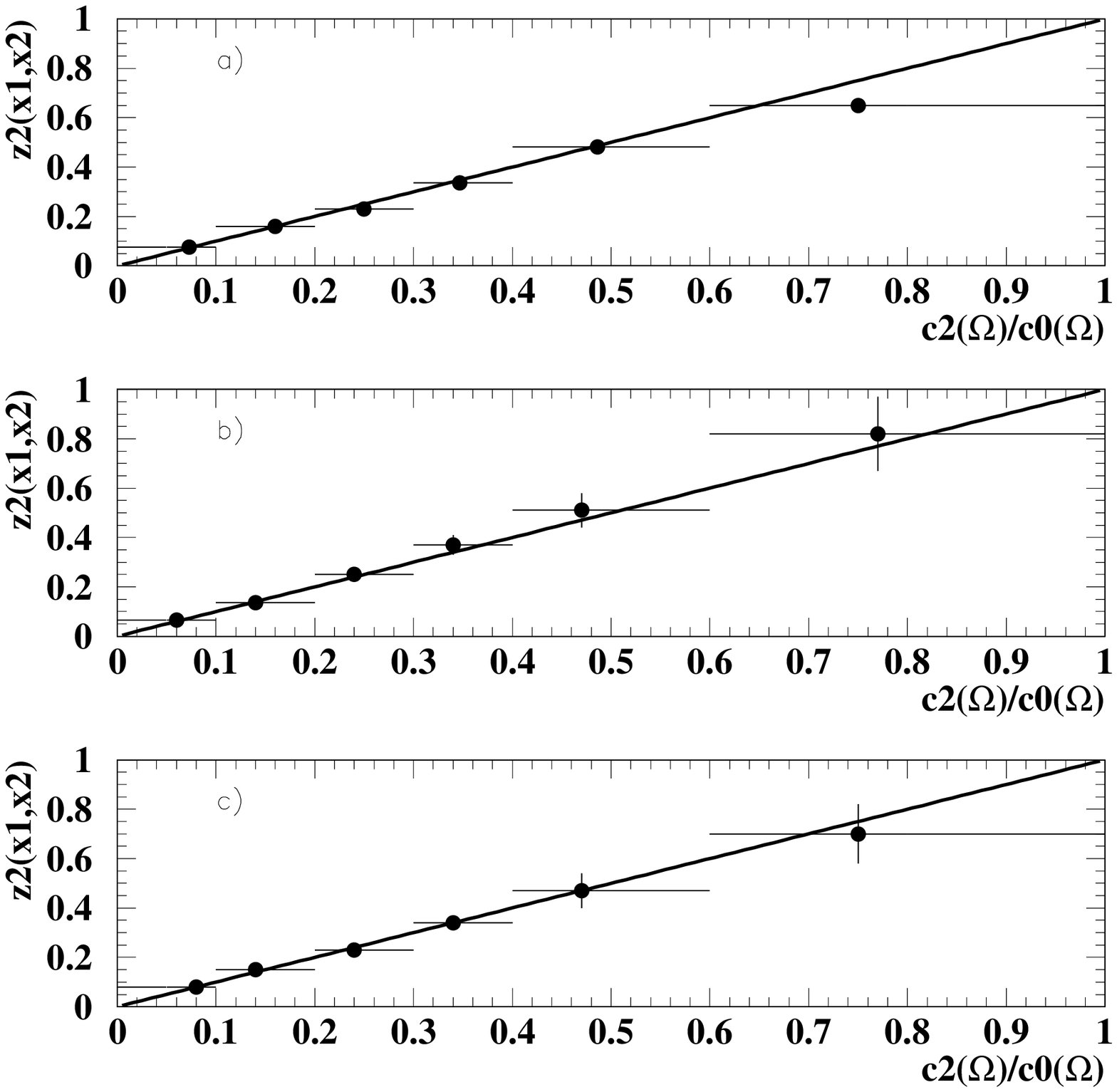,height=18cm}}
\caption{ 
The dependence  of $z_{2}(x_{1},x_{2})$
on the $c_{2}(\vec{\Omega})/c_{0}(\vec{\Omega})$  in three different
regions (a) -2.00--0.020, b) -0.20-0.25 and c) 0.25-1.00) of $c_{1}(\vec{\Omega})/c_{0}(\vec{\Omega})$  when
the $\alpha_{W\phi}$ TGC model is consider for electronic final states. 
The horizontal error-bars
correspond to the size of the bin, whilst the solid line denotes 
the region of equality between the two plotted quantities.}
{
\label{o2resol}}
\end{figure}
\begin{figure}[y1awphi]
\centerline{\epsfig{file=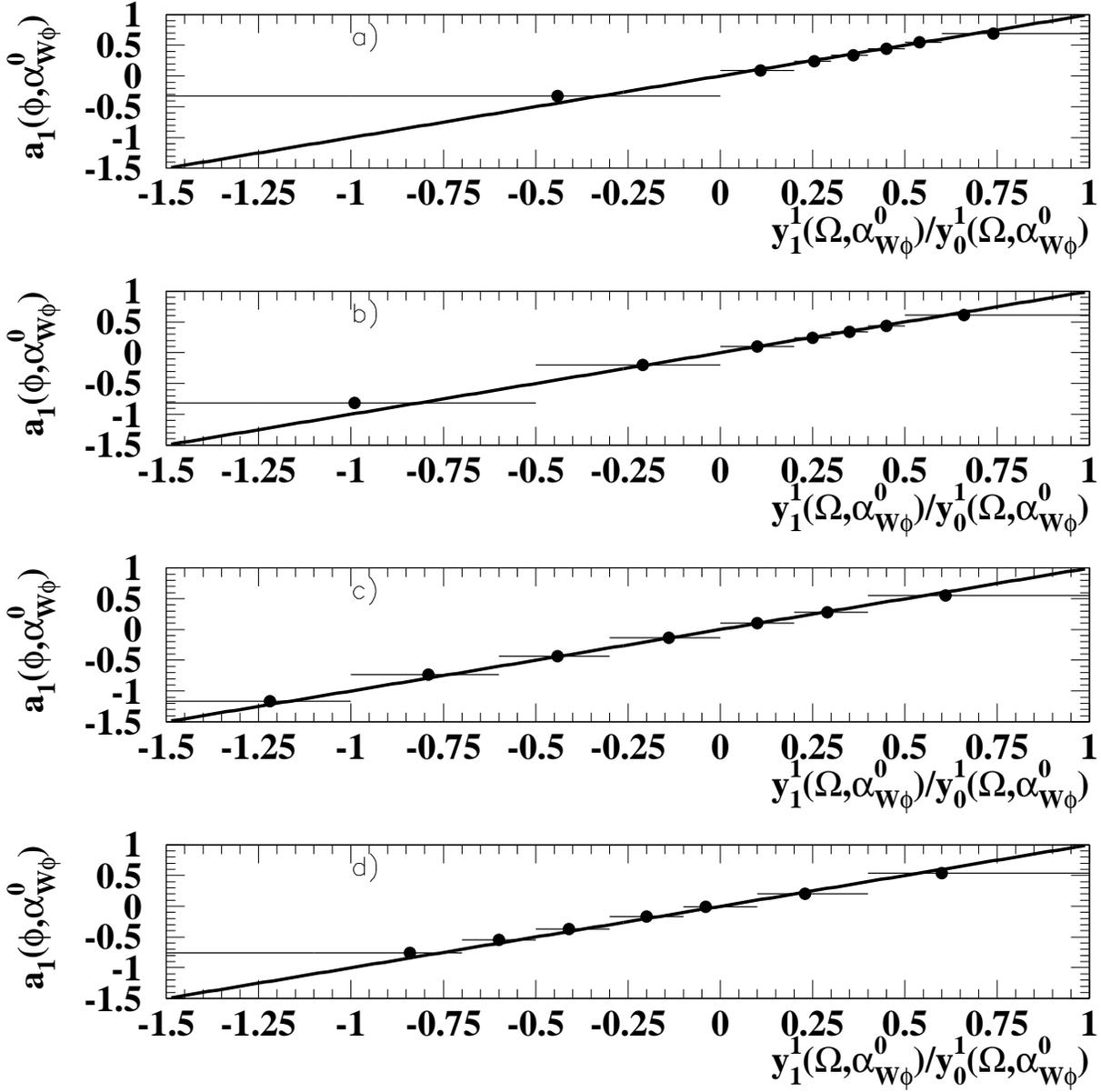,height=18cm}}
\caption{ The approximated Optimum Observable $a_{1}(\phi;\alpha_{W\phi}^{0})$ , corresponding to the
$\alpha_{W\phi}$ coupling for electronic final states, as a function of the quantity 
$y^{1}_{1}(\vec{\Omega};\alpha_{W\phi}^{0})/
y_{0}(\vec{\Omega};\alpha_{W\phi}^{0})$ for several values of $\alpha_{W\Phi}^{0}$ =
a) +1  b) 0.5  )c -0.5 d) -1. The horizontal error-bars
correspond to the size of the bin, whilst the solid line denotes 
the region of equality between the two plotted quantities.}

{
\label{y1awphi}}
\end{figure}
\begin{figure}[liner]
\centerline{\epsfig{file=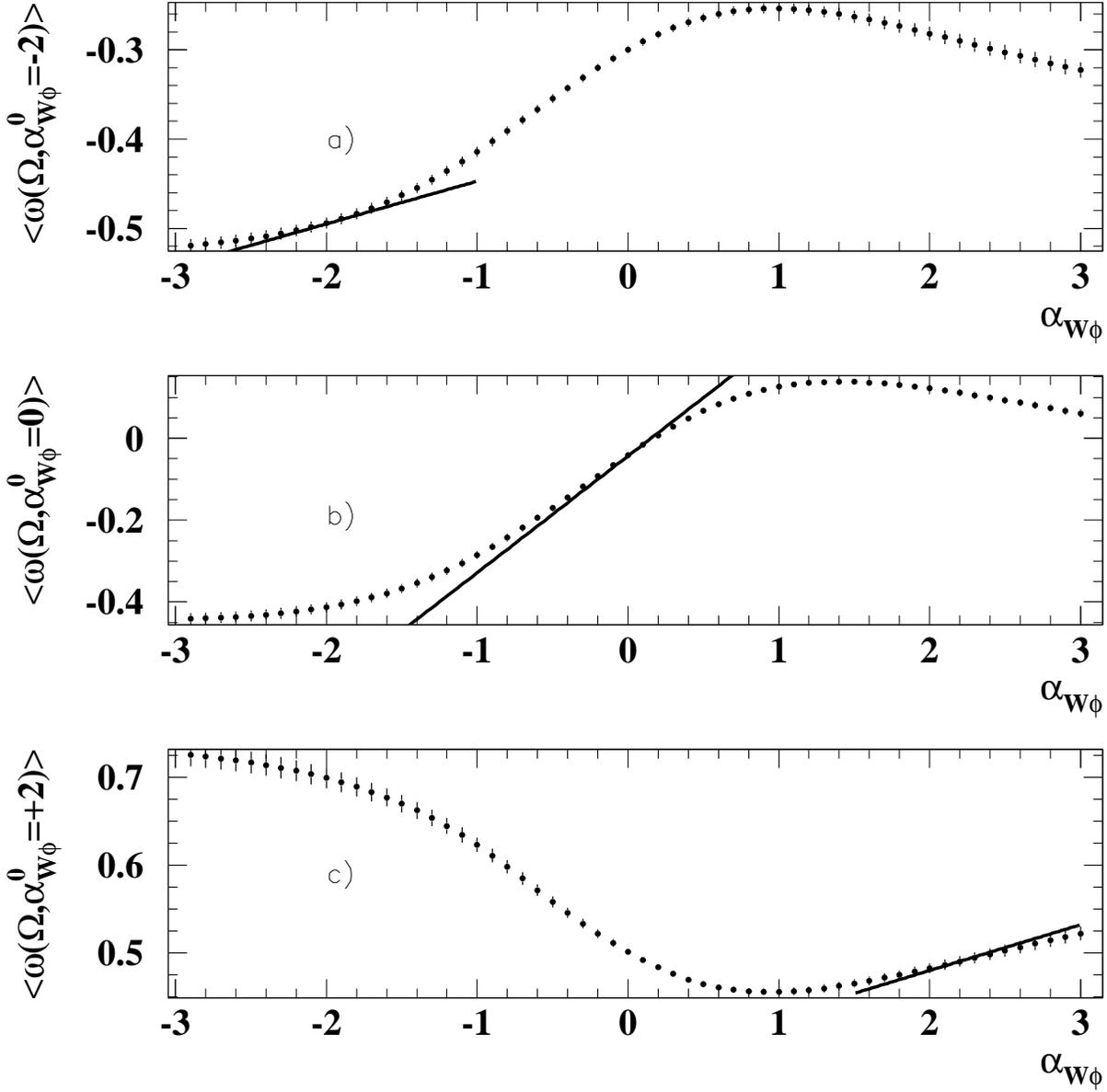,height=18cm}}
\caption{ The mean value of the $\alpha_{W\phi}$  
Modified Observables as a function
of the $\alpha_{W\phi}$  for three different values of the expansion point 
($\alpha_{W\phi}^{0}=-2,0,+2$). The points with the errors correspond to the
reweighted M.C. integration technique  whilst the straight lines represents  the linear equation (51).
}
{
\label{liner}}
\end{figure}
\begin{figure}[consi]
\centerline{\epsfig{file=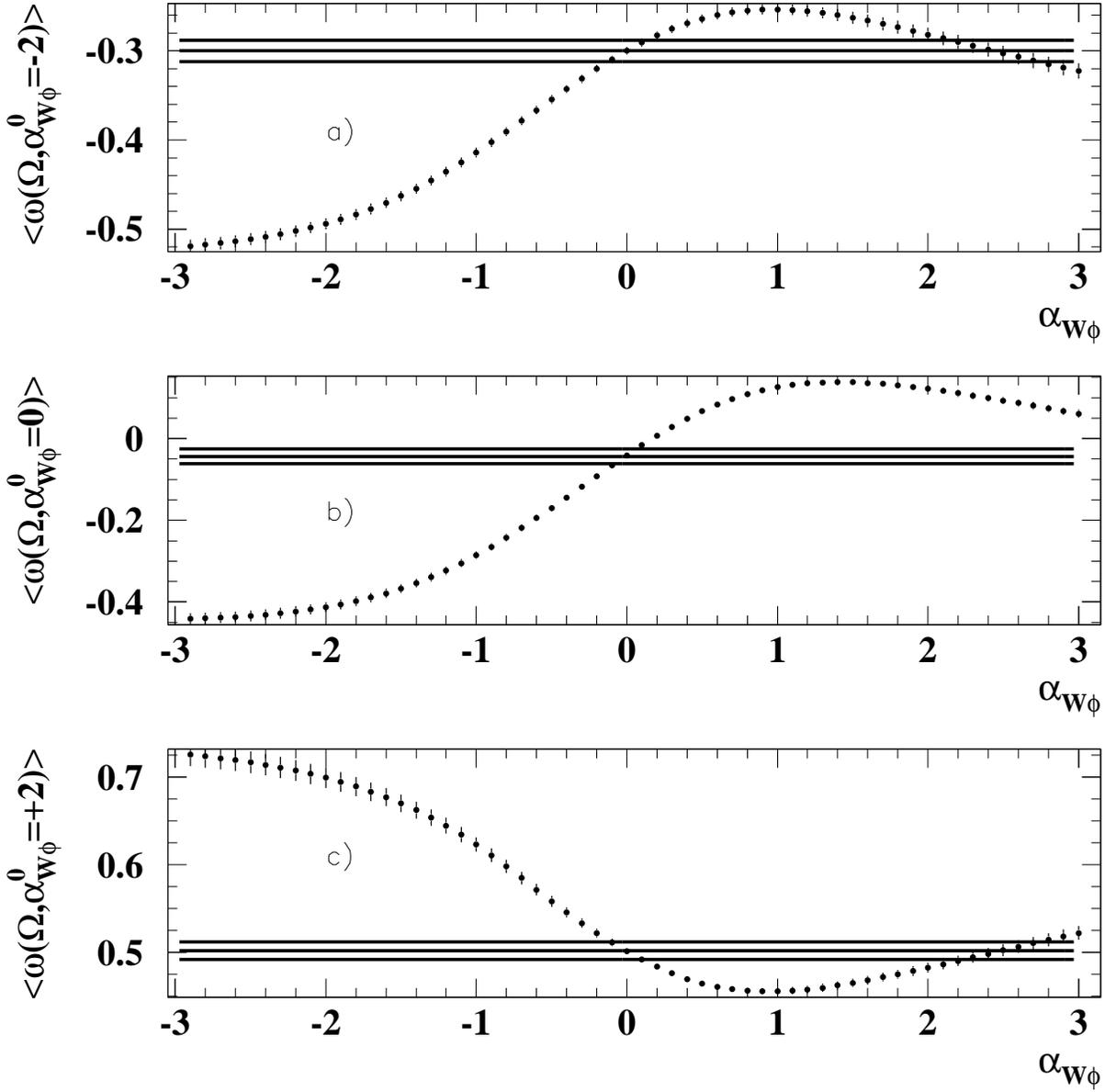,height=18cm}}
\caption{ Estimation of the $\alpha_{W\phi}$ coupling based on The average of the Modified Observables
for three different expansion pints.}
{
\label{consi}}
\end{figure}
\begin{figure}[iter]
\centerline{\epsfig{file=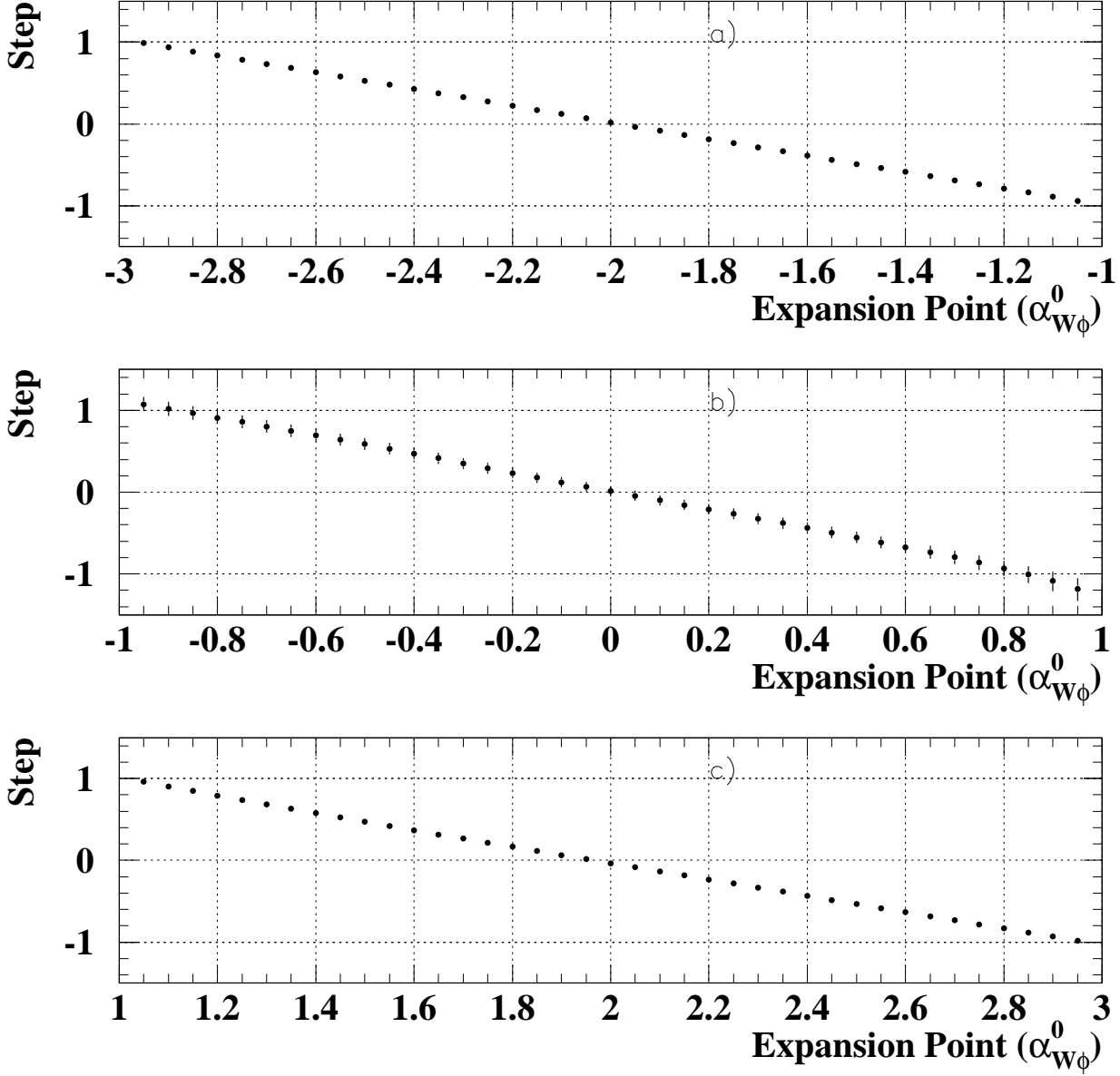,height=18cm}}
\caption{ Estimation of the $\alpha_{W\phi}$ coupling based on the mean value of the Modified Observable.
Three M.C. sets of events produced with different couplings (a) $\alpha_{W\phi}= -2.$,b) $\alpha_{W\phi}= 0.$
and c) $\alpha_{W\phi}= +2.$) were fitted for several values of the expansion point (horizontal axis). The vertical
axis represents the difference of the expansion point from the estimated value in each fit.}
{
\label{iter}}
\end{figure}

\begin{figure}[leiso_awphi]
\centerline{\epsfig{file=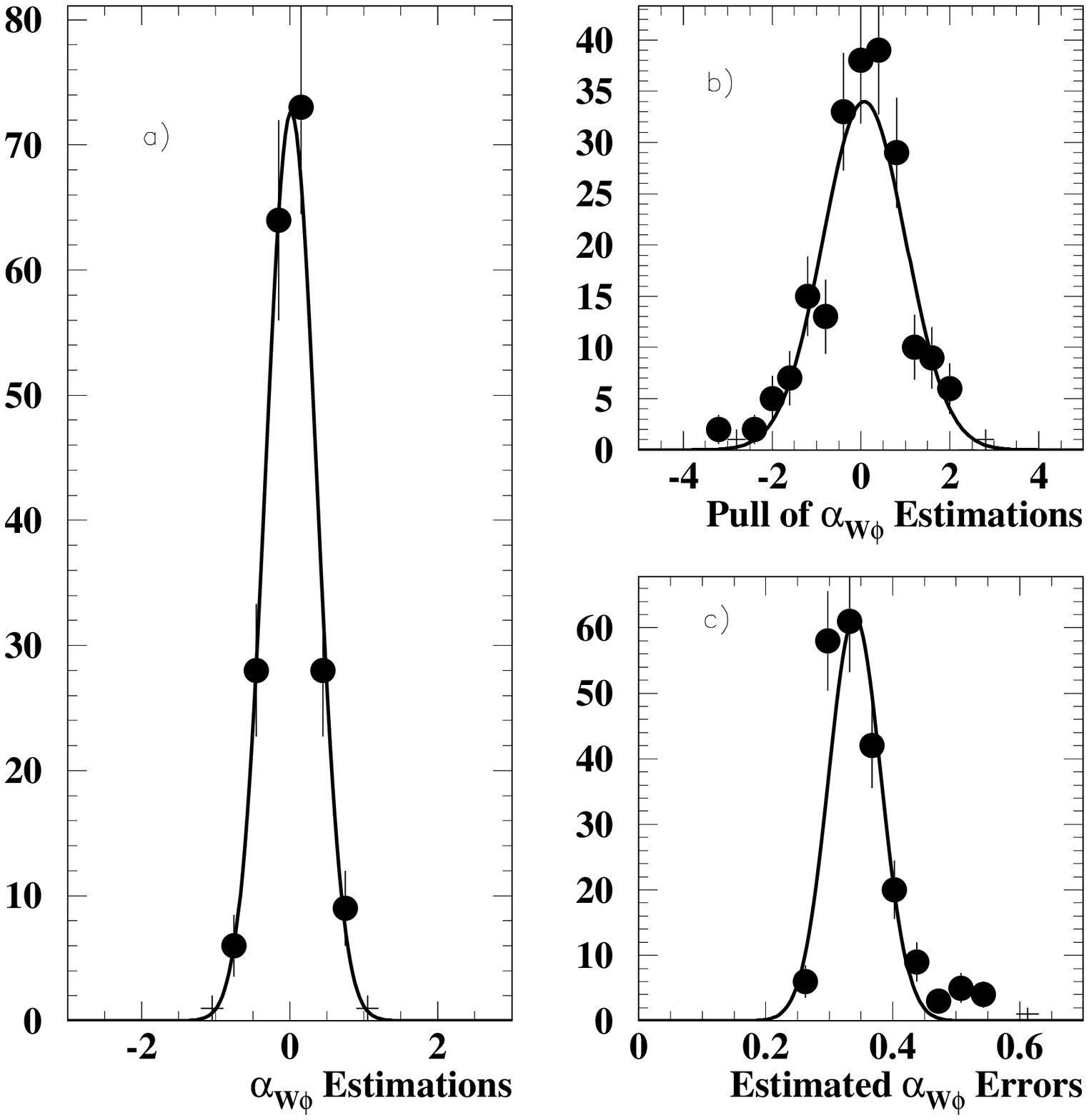,height=18cm}}
\caption{ Results ( a) distribution of the estimations b) pull distribution c) distribution
of the estimated errors) of
Monte Carlo experimentation concerning the estimation of the $\alpha_{W\phi}$ coupling
with binned likelihood fits. The solid lines correspond to gaussian fits.}
{
\label{leisos_awphi}}
\end{figure}

\begin{figure}[leiso_aw]
\centerline{\epsfig{file=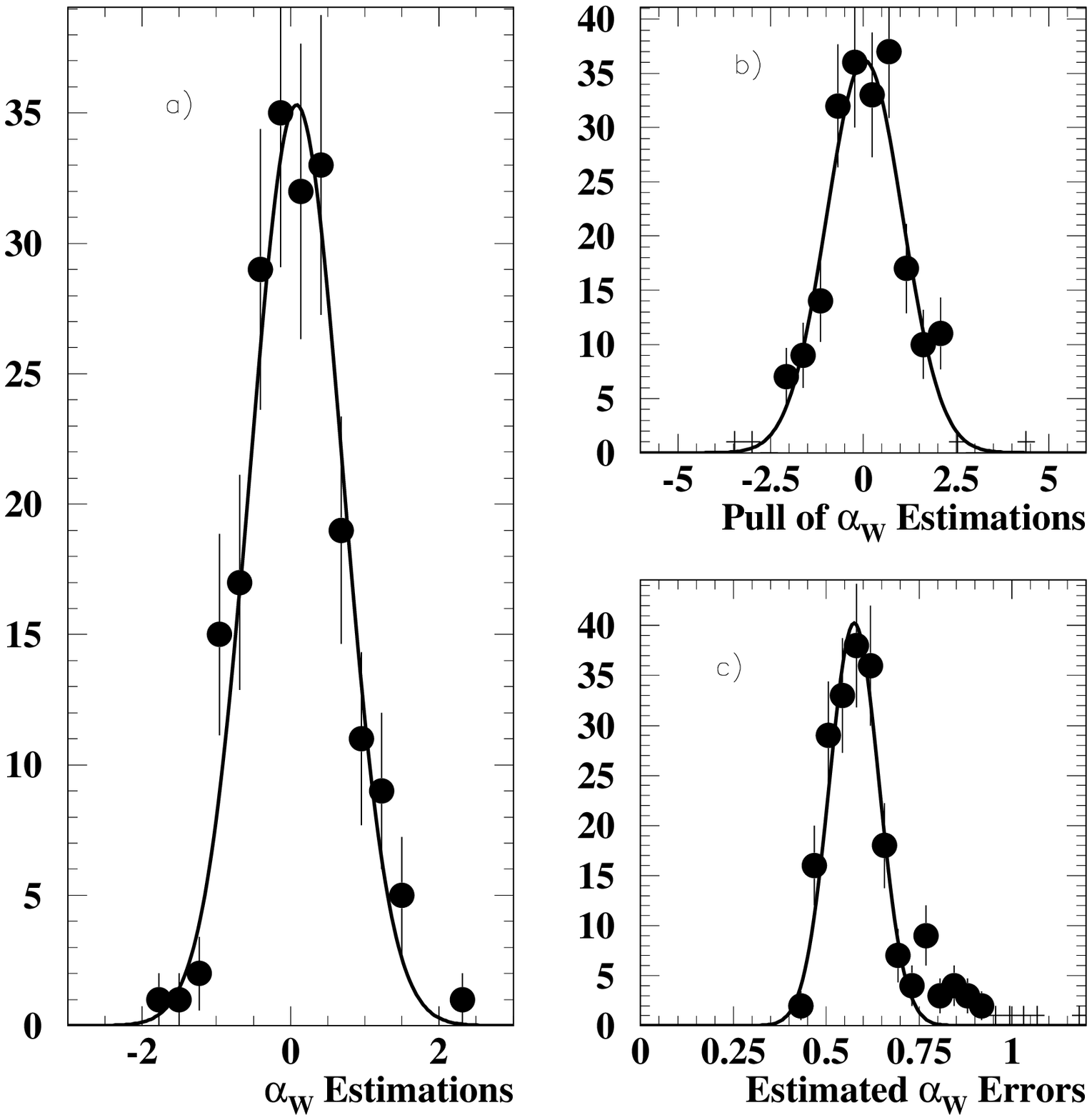,height=18cm}}
\caption{ Results ( a) distribution of the estimations b) pull distribution c) distribution
of the estimated errors) of
Monte Carlo experimentation concerning the estimation of the $\alpha_{W}$ coupling
with binned likelihood fits. The solid lines correspond to gaussian fits.}
{
\label{leisos_aw}}
\end{figure}

\begin{figure}[leiso_bphi]
\centerline{\epsfig{file=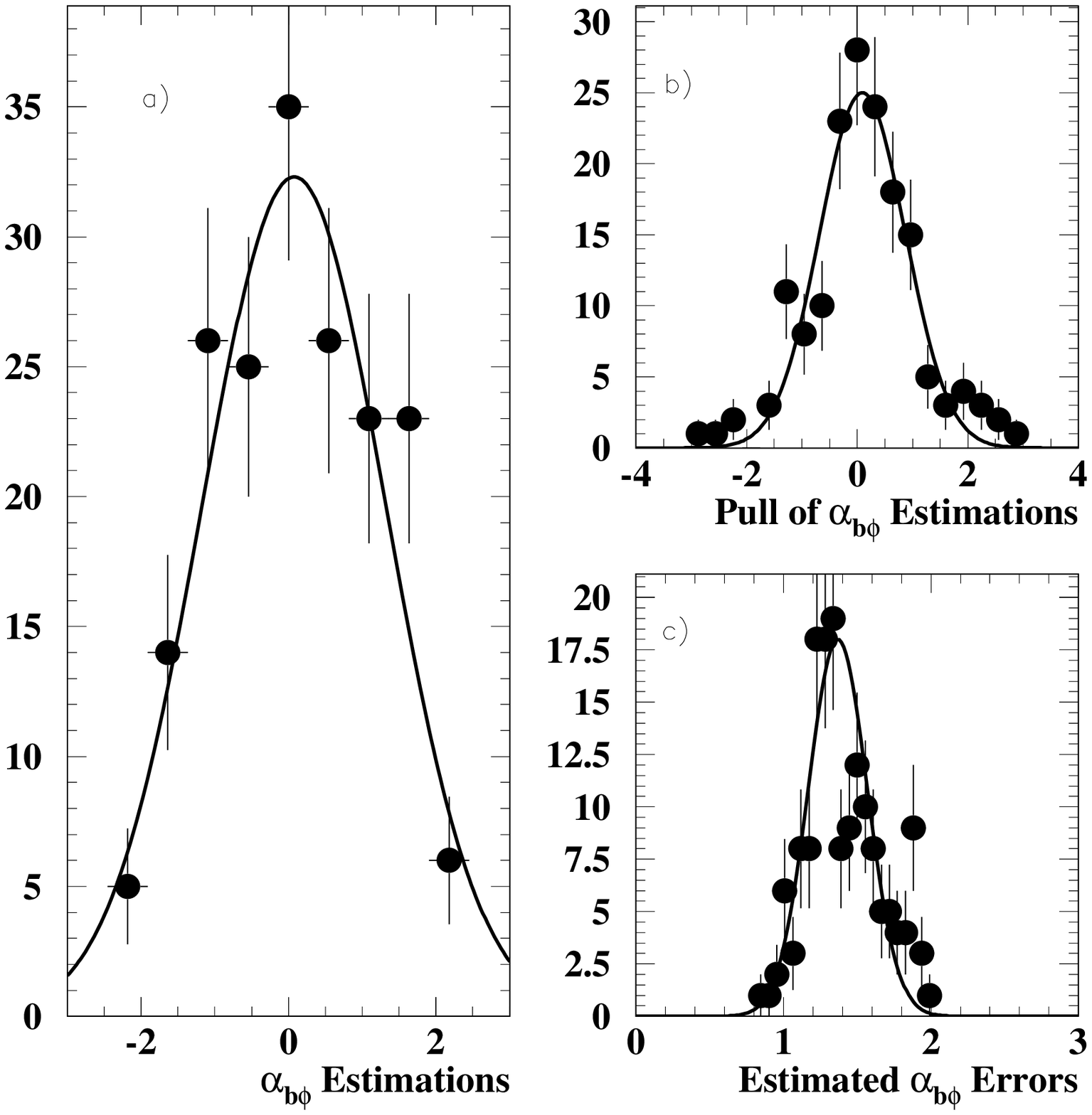,height=18cm}}
\caption{ Results ( a) distribution of the estimations b) pull distribution c) distribution
of the estimated errors) of
Monte Carlo experimentation concerning the estimation of the $\alpha_{B\phi}$ coupling
with binned likelihood fits. The solid lines correspond to gaussian fits.}
{
\label{leisos_bphi}}
\end{figure}

\begin{figure}[sample_awphi]
\centerline{\epsfig{file=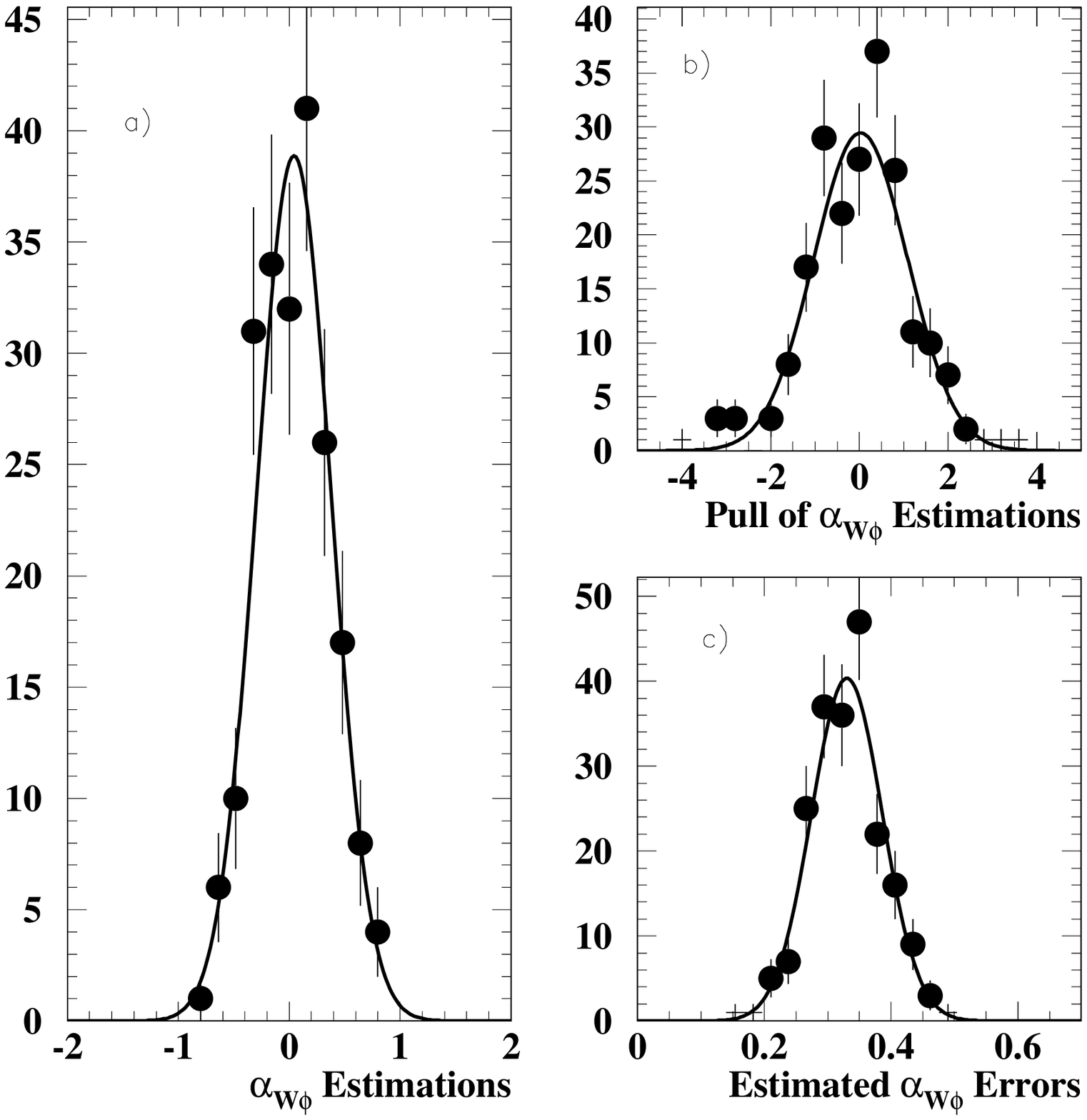,height=18cm}}
\caption{Results ( a) distribution of the estimations b) pull distribution c) distribution
of the estimated errors) of
Monte Carlo experimentation concerning the estimation of the $\alpha_{W\phi}$ coupling
based on the mean value of the Modified Observables. The solid lines correspond to gaussian fits.}
{
\label{sample_awphi}}
\end{figure}

\begin{figure}[sample_aw]
\centerline{\epsfig{file=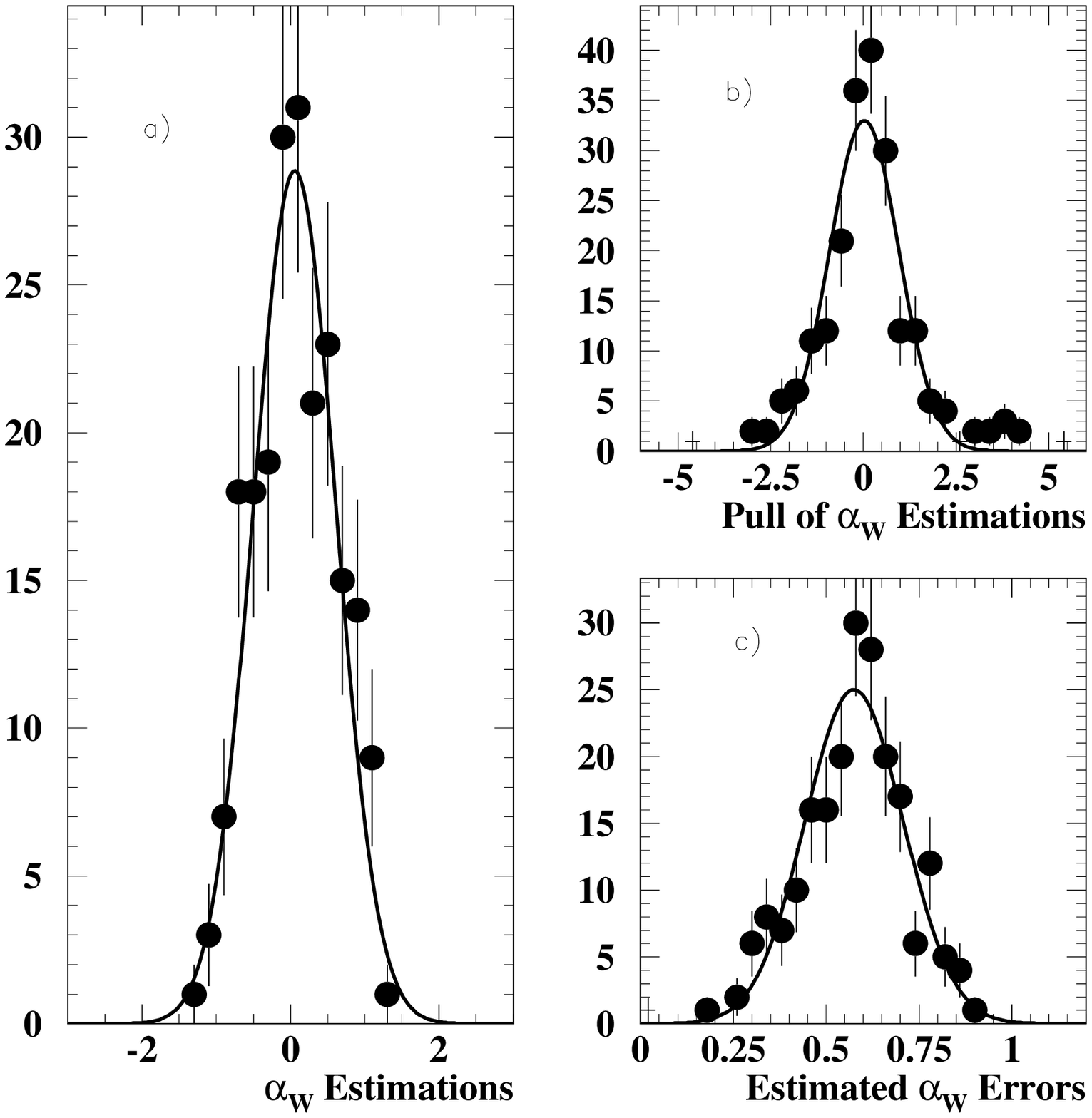,height=18cm}}
\caption{ Results ( a) distribution of the estimations b) pull distribution c) distribution
of the estimated errors) of
Monte Carlo experimentation concerning the estimation of the $\alpha_{W}$ coupling
based on the mean value of the Modified Observables. The solid lines correspond to gaussian fits.}
{
\label{sample_aw}}
\end{figure}

\begin{figure}[sample_bphi]
\centerline{\epsfig{file=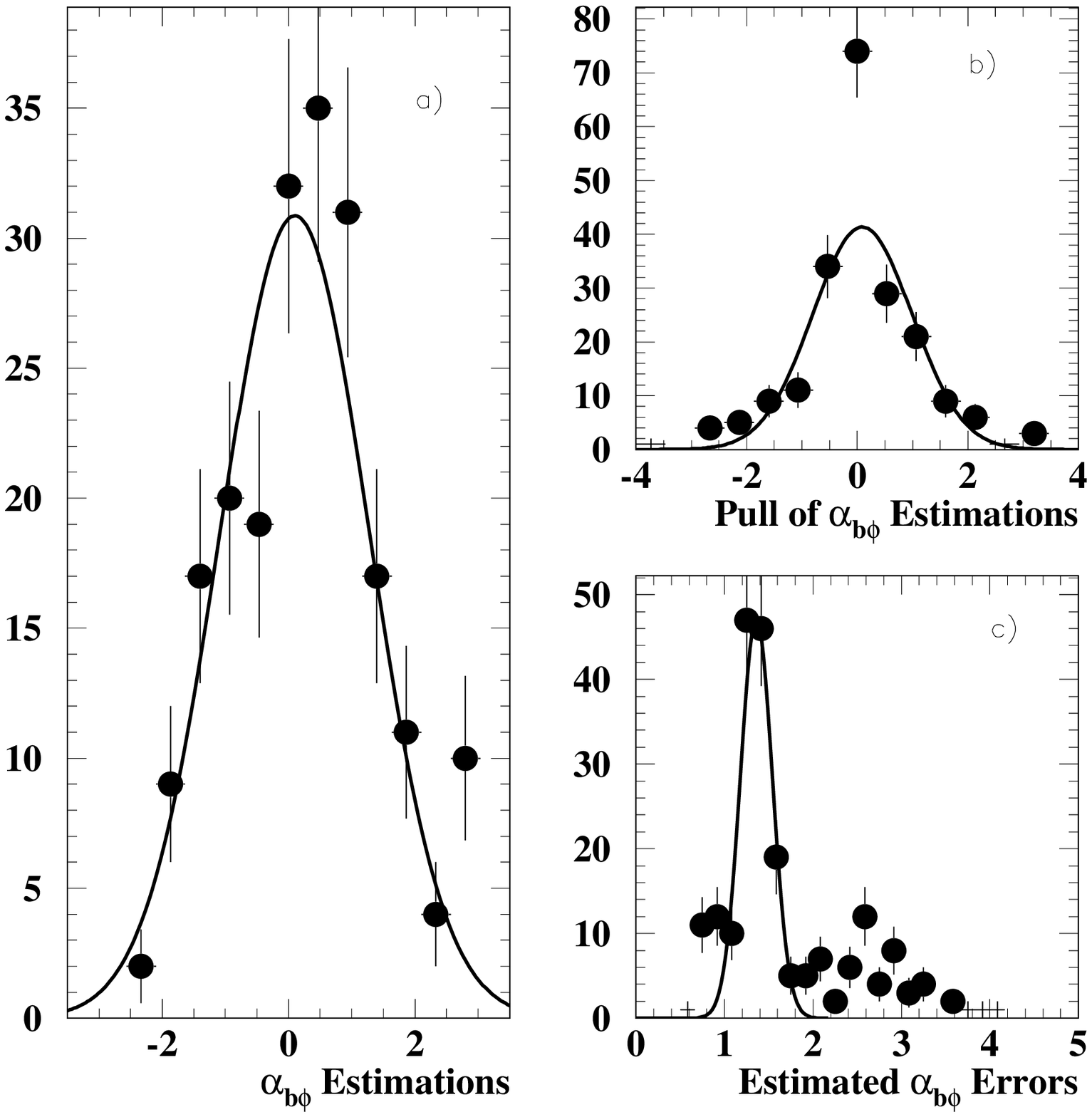,height=18cm}}
\caption{ Results ( a) distribution of the estimations b) pull distribution c) distribution
of the estimated errors) of
Monte Carlo experimentation concerning the estimation of the $\alpha_{B\phi}$ coupling
based on the mean value of the Modified Observables. The solid lines correspond to gaussian fits.}
{
\label{sample_bphi}}
\end{figure}

\vfill
\end{document}